\documentclass[11pt]{article}

\usepackage[american]{babel}
\usepackage{a4}
\usepackage{makeidx}
\usepackage{latexsym}
\usepackage{amsmath}
\usepackage{amsfonts}
\usepackage{mathrsfs}
\usepackage{bm}
\usepackage{multirow}
\usepackage{graphicx}
\usepackage{epsfig}
\usepackage[tight]{subfigure}

\usepackage{enumitem}
\usepackage{times}
\usepackage[compact]{titlesec}

\allowdisplaybreaks  
\renewcommand{\baselinestretch}{1.96}

\usepackage[standard]{ntheorem}

\theoremstyle{nonumberplain}

\theorembodyfont{\upshape} \theoremstyle{plain} \theoremstyle{plain}
\theorembodyfont{\slshape}

\newtheorem{Prop}{Proposition}
\theorembodyfont{\upshape}

\begin{document}
\renewcommand{\baselinestretch}{1.5}
\title{Evolutionary Stochastic Search \\
for Bayesian model exploration}
\author{Leonardo Bottolo \\
Institute for Mathematical Sciences, Imperial College London, UK \\
l.bottolo@imperial.ac.uk \and Sylvia Richardson{\small \thanks{
Address for correspondence: Sylvia Richardson, Department of
Epidemiology and Public Health, Imperial College, 1 Norfolk Place,
London, W2 1PG, UK.}} \\
Centre for Biostatistics, Imperial College, London, UK \\
sylvia.richardson@imperial.ac.uk}
\date{}
\maketitle
\renewcommand{\baselinestretch}{2}

\begin{abstract}
Implementing Bayesian variable selection for linear Gaussian
regression models for analysing high dimensional data sets is of
current interest in many fields. In order to make such analysis
operational, we propose a new sampling algorithm based upon
Evolutionary Monte Carlo and designed to work under the \lq\lq large
$p$, small $n$\rq\rq\ paradigm, thus making fully Bayesian
multivariate analysis feasible, for example, in genetics/genomics
experiments. Two real data examples in genomics are presented,
demonstrating the performance of the algorithm in a space of up to
$10,000$ covariates. Finally the methodology is compared with a
recently proposed search algorithms in an extensive simulation
study.
\end{abstract}


\clearpage

\section{Introduction}
This paper is a contribution to the methodology of Bayesian variable
selection for linear Gaussian regression models, an important
problem which has been much discussed both from a theoretical and a
practical perspective (see Chipman \emph{et al.}, 2001 and Clyde and
George, 2004 for literature reviews). Recent advances have been made
in two directions, unravelling the theoretical properties of
different choices of prior structure for the regression coefficients
(Fern\'{a}ndez \emph{et al.}, 2001; Liang \emph{et al.}, 2008) and
proposing algorithms that can explore the huge model space
consisting of all the possible subsets when there are a large number
of covariates, using either MCMC or other search algorithms (Kohn
\emph{et al.}, 2001; Dellaportas \emph{et al.}, 2002; Hans \emph{et
al.}, 2007).

In this paper, we propose a new sampling algorithm for implementing
the variable selection model, based on tailoring ideas from
Evolutionary Monte Carlo (Liang and Wong, 2000; Jasra \emph{et al.},
2007; Wilson \emph{et al.}, 2009) in order to overcome the known
difficulties that MCMC samplers face in a high dimension multimodal
model space: enumerating the model space becomes rapidly unfeasible
even for a moderate number of covariates. For a Bayesian approach to
be operational, it needs to be accompanied by an algorithm that
samples the indicators of the selected subsets of covariates,
together with any other parameters that have not been integrated
out. Our new algorithm for searching through the model space has
many generic features that are of interest \emph{per se} and can be
easily coupled with any prior formulation for the
variance-covariance of the regression coefficients. We illustrate
this by implementing $g$-priors for the regression coefficients as
well as independent priors: in both cases the formulation we adopt
is general and allows the specification of a further level of
hierarchy on the priors for the regression coefficients, if so
desired.

The paper is structured as follows. In Section \ref{Background}, we
present the background of Bayesian variable selection, reviewing
briefly alternative prior specifications for the regression
coefficients, namely $g$-priors and independent priors. Section
\ref{MCMC sampler} is devoted to the description of our MCMC sampler
which uses a wide portfolio of moves, including two proposed new
ones. Section \ref{Performance_ESS} demonstrates the good
performance of our new MCMC algorithm in a variety of real and
simulated examples with different structures on the predictors. In
Section \ref{Simulation study} we complement the results of the
simulation study by comparing our algorithm with the recent Shotgun
Stochastic Search algorithm of Hans \emph{et al.} (2007). Finally
Section \ref{Discussion} contains some concluding remarks.

\section{Background} \label{Background}
\subsection{Variable selection} \label{Variable selection}
\noindent Let $y=\left( y_{1},\ldots ,y_{n}\right) ^{T}$  be a
sequence of $n$ observed responses and $x_{i}=\left( x_{i1},\ldots
,x_{ip}\right) ^{T}$ a vector of predictors for $y_{i}$, $i=1,\ldots
,n$, of dimension $p\times 1$. Moreover let $X$ be the $n\times p$
design matrix with $i$th row $x_{i}^{T}$. A Gaussian linear model
can be described by the equation
\[
y=\alpha 1_{n}+X\beta +\varepsilon,
\]
where $\alpha $ is an unknown constant,  $1_{n}$ is a column vector
of ones, $\beta =\left( \beta _{1},\ldots ,\beta _{p}\right) ^{T}$
is a $p\times 1$ vector of unknown parameters and $\varepsilon \sim
N\left( 0,\sigma ^{2}I_{n}\right) $.

Suppose one wants to model the relationship between $y$ and a subset
of $x_{1},\ldots ,x_{p}$, but there is uncertainty about which
subset to use. Following the usual convention of only considering
models that have the intercept $\alpha $, this problem, known as
variable selection or subset selection, is particularly interesting
when $p$ is large and parsimonious models containing only a few
predictors are sought to gain interpretability. From a Bayesian
perspective the problem is tackled by placing a constant prior
density on $\alpha $ and a prior on $\beta $ which depends on a
latent binary vector $\gamma =\left( \gamma _{1},\ldots ,\gamma
_{p}\right) ^{T}$, where $\gamma _{j}=1$ if $\beta _{j}\neq 0$ and
$\gamma _{j}=0$ if $\beta _{j}=0$, $j=1,\ldots ,p$. The overall
number of possible models defined through $\gamma$ grows
exponentially with $p$ and selecting the best model that predicts
$y$ is equivalent to find one over the $2^{p}$ subsets that form the
model space.

Given the latent variable $\gamma $, a Gaussian linear model can
therefore be written as
\begin{equation}
y=\alpha 1_{n}+X_{\gamma }\beta _{\gamma }+\varepsilon , \label{T1}
\end{equation}
where $\beta _{\gamma }$ is the non-zero vector of coefficients
extracted from $\beta $, $X_{\gamma }$ is the design matrix of
dimension $n\times p_{\gamma }$, $p_{\gamma }\equiv \gamma^{T}
1_{p}$, with columns corresponding to $\gamma _{j}=1$. We will
assume that, apart from the intercept $\alpha $, $x_{1},\ldots
,x_{p}$ contains no variables that would be included in every
possible model and that the columns of the design matrix have all
been centred with mean $0$.

It is recommended to treat the intercept separately and assign it a
constant prior: $p\left( \alpha \right) \propto 1$, Fern\'{a}ndez
\emph{et al.} (2001). When coupled with the latent variable $\gamma
$, the conjugate prior structure of $\left( \beta _{\gamma },\sigma
^{2}\right) $ follows a normal-inverse-gamma distribution
\begin{equation}
p\left( \beta _{\gamma }\left\vert \gamma ,\sigma ^{2}\right.
\right) =N\left( m_{\gamma },\sigma ^{2}\Sigma _{\gamma } \right)
\label{T2}
\end{equation}
\begin{equation}
p\left( \sigma ^{2}\left\vert \gamma \right. \right) =p\left( \sigma
^{2}\right) =InvGa\left( a_{\sigma },b_{\sigma }\right) \label{T3}
\end{equation}
with $a_{\sigma },b_{\sigma }>0$. Some guidelines on how to fix the
value of the hyperparameters $a_{\sigma }$ and $b_{\sigma }$ are
provided in Kohn \emph{et al.} (2001), while the case $a_{\sigma
}=b_{\sigma }=0$ corresponds to the Jeffreys' prior for the error
variance, $p\left( \sigma ^{2}\right) \propto \sigma ^{-2}$. Taking
into account (\ref{T1}), (\ref{T2}), (\ref{T3}) and the prior
specification for $\alpha $, the joint distribution of all the
variables (based on further conditional independence conditions) can
be written as
\begin{equation}
p\left( y,\gamma ,\alpha ,\beta _{\gamma },\sigma ^{2}\right)
=p\left( y\left\vert \gamma ,\alpha ,\beta _{\gamma },\sigma
^{2}\right. \right) p\left( \alpha \right) p\left( \beta _{\gamma
}\left\vert \gamma ,\sigma ^{2}\right. \right) p\left( \sigma
^{2}\right) p\left( \gamma \right). \label{T4}
\end{equation}

The main advantage of the conjugate structure (\ref{T2}) and
(\ref{T3}) is the analytical tractability of the marginal likelihood
whatever the specification of the prior covariance matrix $\Sigma
_{\gamma }$:
\begin{eqnarray}
&&\int p\left( y\left\vert \gamma ,\alpha ,\beta _{\gamma },\sigma
^{2}\right. \right) p\left( \alpha \right) p\left( \beta _{\gamma
}\left\vert \gamma ,\sigma ^{2}\right. \right) p\left( \sigma
^{2}\right)
d\alpha \text{\/{}\/}d\beta _{\gamma }d\sigma ^{2}  \notag \\
&\propto &\left\vert X_{\gamma }^{T}X_{\gamma }+\Sigma _{\gamma
}^{-1}\right\vert ^{-1/2}\left\vert \Sigma _{\gamma }\right\vert
^{-1/2}\left( 2b_{\sigma }+S\left( \gamma \right) \right) ^{-\left(
2a_{\sigma }+n-1\right) /2},  \label{T5}
\end{eqnarray}
where $S\left( \gamma \right) =C-M^{T}K_{\gamma }^{-1}M$, with
$C=\left( y-\bar{y}_{n}\right) ^{T}\left( y-\bar{y}_{n}\right)
+m_{\gamma }^{T}\Sigma _{\gamma }^{-1}m_{\gamma }$, $M=X_{\gamma
}^{T}\left( y-\bar{y}_{n}\right)+\Sigma _{\gamma }^{-1}m_{\gamma }$
and $K_{\gamma }=X_{\gamma }^{T}X_{\gamma }+\Sigma _{\gamma }^{-1}$
(Brown\emph{ et al.}, 1998).

\noindent While the mean of the prior (\ref{T2}) is usually set
equal to zero, $m_{\gamma }=0$, a neutral choice (Chipman \emph{et
al.}, 2001; Clyde and George, 2004), the specification of the prior
covariance $\Sigma _{\gamma }$ matrix leads to at least two
different classes of priors:
\begin{itemize}[leftmargin=*, itemsep=-0.5em]
\item When $\Sigma _{\gamma }=gV_{\gamma }$, where $g$ is a scalar
  and $V_{\gamma }=\left( X_{\gamma }^{T}X_{\gamma }\right) ^{-1}$,
  it replicates the covariance structure of the likelihood giving
  rise to so called $g$-priors first proposed by Zellner (1986).
\item When $\Sigma _{\gamma }=cV_{\gamma }$, but $V_{\gamma }=I_{p_{\gamma
  }}$ the components of $\beta _{\gamma }$ are conditionally
  independent and the posterior covariance matrix is driven towards
  the independence case.
\end{itemize}
We will adopt the notation $\Sigma _{\gamma }=\tau V_{\gamma }$ as
we want to cover both prior specification in a unified manner. Thus
in the $g$-prior case, $\Sigma _{\gamma }=\tau\left( X_{\gamma
}^{T}X_{\gamma }\right) ^{-1}$ while in the independent case,
$\Sigma _{\gamma }=\tau I_{p_{\gamma}}$. We will refer to $\tau$ as
the \emph{variable selection coefficient} for reasons that will
become clear in the next Section.

To complete the prior specification in (\ref{T4}), $p\left( \gamma
\right) $ must be defined. A complete discussion about alternative
priors on the model space can be found in Chipman (1996) and Chipman
\emph{et al.} (2001). Here we adopt the beta-binomial prior
illustrated in Kohn \emph{et al.} (2001)
\begin{equation}
p\left( \gamma \right) =\int p\left( \gamma \left\vert \omega
\right. \right) p\left( \omega \right) d\omega =\frac{B\left(
p_{\gamma }+a_{\omega },p-p_{\gamma }+b_{\omega }\right) }{B\left(
a_{\omega },b_{\omega }\right) } \label{T7}
\end{equation}
with $p_{\gamma }\equiv \gamma^{T} 1_{p}$, where the choice $p\left(
\gamma \left\vert \omega \right. \right) =\omega ^{p_{\gamma
}}\left( 1-\omega \right) ^{p-p_{\gamma }}$ implicitly induces a
binomial prior distribution over the model size and $p\left( \omega
\right) =\omega ^{a_{\omega }-1}\left( 1-\omega \right) ^{b_{\omega
}-1}/B\left( a_{\omega },b_{\omega }\right) $. The hypercoefficients
$a_{\omega }$ and $b_{\omega }$ can be chosen once $E\left(p_{\gamma
}\right) $ and $V\left( p_{\gamma }\right) $ have been elicited. In
the \lq\lq large $p$, small $n$\rq\rq\ framework, to ensure sparse
regression models where $p_{\gamma }\ll p$, it is recommended to
centre the prior for the model size away from the number of
observations.

\subsection{Priors for the variable selection coefficient $\tau$} \label{SelectionCoeff}
\subsubsection{$g$-priors} \label{g-priors}
It is a known fact that $g$-priors have two attractive properties.
Firstly they possess an automatic scaling feature (Chipman \emph{et
al.}, 2001; Kohn \emph{et al.}, 2001). In contrast, for independent
priors, the effect of $V_{\gamma }=I_{p_{\gamma }}$ on the posterior
distribution depends on the relative scale of $X$ and
standardisation of the design matrix to units of standard deviation
is recommended. However, this is not always the best procedure when
$X$ is possibly skewed, or when the columns of $X$ are not defined
on a common scale of measurement. The second feature that makes
$g$-priors particularly appealing is the rather simple structure of
the marginal likelihood (\ref{T5}) with respect to the constant
$\tau$ which becomes
\begin{equation}
\propto \left( 1+\tau\right) ^{-p_{\gamma }/2}\left( 2b_{\sigma
}+S\left( \gamma \right) \right) ^{-\left( 2a_{\sigma }+n-1\right)
/2}, \label{T6}
\end{equation}
where, if $m_{\gamma }=0$, $S\left( \gamma \right) =\left(
y-\bar{y}_{n}\right) ^{T}\left( y-\bar{y} _{n}\right) -\frac{\tau
}{1+\tau }\left( y-\bar{y}_{n}\right) ^{T}X_{\gamma }\left(
X_{\gamma }^{T}X_{\gamma }\right) ^{-1}X_{\gamma }^{T}\left(
y-\bar{y }_{n}\right) $. For computational reasons explained in the
next Section, we assume that (\ref{T6}) is always defined: since we
calculate $S\left( \gamma \right)$ using the QR-decomposition of the
regression $\left( X_{\gamma },y-\bar{y}_{n}\right) $ (Brown
\emph{et al.}, 1998), when $n\leq p_{\gamma }$, $S\left( \gamma
\right) =\left( y-\bar{y}_{n}\right) ^{T}\left( y-\bar{y}
_{n}\right) /\left( 1+\tau \right) $. Despite the simplicity of
(\ref{T6}), the choice of the constant $\tau$ for $g$-priors is
complex, see Fern\'{a}ndez \emph{et al.} (2001), Cui and George
(2008) and Liang \emph{et al.} (2008).

Historically the first attempt to build a comprehensive Bayesian
analysis placing a prior distribution on $\tau$ dates back to
Zellner and Siow (1980), where the data adaptivity of the degree of
shrinkage adapts to different scenarios better than assuming
standard fixed values. Zellner-Siow priors, Z-S hereafter, can be
thought as a mixture of $g$-priors and an inverse-gamma prior on
$\tau$, $\tau \sim InvGa(1/2, n/2)$, leading to
\begin{equation}
p\left( \beta _{\gamma }\left\vert \gamma ,\sigma ^{2}\right.
\right) \propto \int N\left( 0,\sigma ^{2}\tau \left( X_{\gamma
}^{T}X_{\gamma }\right) ^{-1}\right) p\left( \tau\right) d\tau.
\label{T9}
\end{equation}
Liang \emph{et al.} (2008) analyse in details Z-S priors pointing
out a variety of theoretical properties. From a computational point
of view, with Z-S priors, the marginal likelihood $p\left(
y\left\vert \gamma \right. \right) = \int p\left( y\left\vert \gamma
,\tau \right. \right) p\left( \tau \right) d\tau$ is no more
available in closed form, something which is advantageous in order
to quickly perform a stochastic search (Chipman \emph{et al.},
2001). Even though Z-S priors need no calibration and the Laplace
approximation can be derived (Tierney and Kadane, 1986), see
Appendix \ref{Laplace_approx}, never became as popular as $g$-priors
with a suitable constant value for $\tau$. For alternative priors,
see also Cui and George (2008) and Liang \emph{et al.} (2008).

\subsubsection{Independent priors} \label{Independent priors}
When all the variables are defined on the same scale, independent
priors represent an attractive alternative to $g$-priors. The
likelihood marginalised over $\alpha$, $\beta _{\gamma }$ and
$\sigma ^{2}$ becomes
\begin{eqnarray}
p\left( y\left\vert \gamma \right. \right) \propto \tau^{-p_{\gamma
}/2}\left\vert X_{\gamma }^{T}X_{\gamma }+ \tau I_{p_{\gamma
}}\right\vert ^{-1/2}\left( 2b_{\sigma }+S\left( \gamma \right)
\right) ^{-\left( 2a_{\sigma }+n-1\right) /2}, \label{T13}
\end{eqnarray}
where, if $m_{\gamma }=0$, $S\left( \gamma \right) =\left(
y-\bar{y}_{n}\right) ^{T}\left( y-\bar{y}_{n}\right) -\left(
y-\bar{y}_{n}\right) ^{T}X_{\gamma }\left( X_{\gamma }^{T}X_{\gamma
}+ \tau I_{p_{\gamma }}\right) ^{-1}X_{\gamma }^{T}\left(
y-\bar{y}_{n}\right) $. Note that (\ref{T13}) is computationally
more demanding than (\ref{T6}) due to the extra determinant
operator.

Geweke (1996) suggests to fix a different value of $\tau_{j}$,
$j=1,\ldots ,p$, based on the idea of \lq\lq substantially
significant determinant\rq\rq\ of $\Delta X_{j}$ with respect to
$\Delta y$. However it is common practice to standardise the
predictor variables, taking $\tau=1$ in order to place appropriate
prior mass on reasonable values of the regression coefficients (Hans
\emph{et al.}, 2007). Another approach, illustrated in Bae and
Mallick (2004), places a prior distribution on $\tau_{j}$ without
standardising the predictors.

Regardless of the prior specification for $\tau$, using the
QR-decomposition on a suitable transformation of $ X_{\gamma }$ and
$y-\bar{y}_{n} $, the marginal likelihood (\ref{T13}) is always
defined.

\section{MCMC sampler} \label{MCMC sampler}
In this Section we propose a new sampling algorithm that overcomes
the known difficulties faced by MCMC schemes when attempting to
sample a high dimension multimodal space. We discuss in a unified
manner the general case where a hyperprior on the variable selection
coefficient $\tau$ is specified. This encompasses the $g$-prior and
independent prior structure as well as the case of fixed $\tau$ if a
point mass prior is used.

The multimodality of the model space is a known issue in variable
selection and several ways to tackle this problem have been proposed
in the past few years. Liang and Wong (2000) suggest an extension of
parallel tempering called Evolutionary Monte Carlo, EMC hereafter,
Nott and Green, N\&G hereafter, (2004) introduce a sampling scheme
inspired by the Swendsen-Wang algorithm while Jasra \emph{et al.}
(2007) extend EMC methods to varying dimension algorithms. Finally
Hans \emph{et al.} (2007) propose when $p>n$ a new stochastic search
algorithm, SSS, to explore models that are in the same neighbourhood
in order to quickly find the best combination of predictors.

We propose to solve the issue related to the multimodality of model
space (and the dependence between $\gamma $ and $\tau$) along the
lines of EMC, applying some suitable parallel tempering strategies
directly on $p\left( y\left\vert \gamma ,\tau \right. \right) $. The
basic idea of parallel tempering, PT hereafter, is to weaken the
dependence of a function from its parameters by adding an extra one
called \lq\lq temperature\rq\rq. Multiple Markov chains, called
\lq\lq population\rq\rq\ of chains, are run in parallel, where a
different temperature is attached to each chain, their state is
tentatively swap at every sweep by a probabilistic mechanism and the
latent binary vector $\gamma $ of the non-heated chain is recorded.
The different temperatures have the effect of flatting the
likelihood. This ensures that the posterior distribution is not
trapped in any local mode and that the algorithm mixes efficiently,
since every chain constantly tries to transmit information about its
state to the others. EMC extents this idea, encompassing the
positive features of PT and genetic algorithms inside a MCMC scheme.

Since $\beta $ and $\sigma ^{2}$ are integrated out, only two
parameters need to be sampled, namely the latent binary vector and
the variable selection coefficient. In this set-up the full
conditionals to be considered are
\begin{equation}
\left[ p\left( \gamma _{l}\left\vert \cdots \right. \right) \right]
^{1/t_{l}}\propto \left[ p\left( y\left\vert \gamma _{l},\tau
\right. \right) \right] ^{1/t_{l}}\left[ p\left( \gamma _{l}\right)
\right] ^{1/t_{l}} \label{T23bis}
\end{equation}
\begin{equation}
p\left( \tau \left\vert \cdots \right. \right) \propto
\prod\nolimits_{l=1}^{L}\left[ p\left( y\left\vert \gamma _{l},\tau
\right. \right) \right] ^{1/t_{l}}p\left( \tau \right),
\label{T24bis}
\end{equation}
where $L$ is the number of chains in the the population and $t_{l}$,
$1=t_{1}<t_{2}<\cdots <t_{L}$, is the temperature attached to the
$l$th chain while the population $\bm{\gamma }$ corresponds to a set
of chains that are retained simultaneously. Conditions for
convergence of EMC algorithms are well understood and illustrated
for instance in Jasra \emph{et al.} (2007).

At each sweep of our algorithm, first the population $\bm{\gamma }$
in (\ref{T23bis}) is updated using a variety of moves inspired by
genetic algorithms: \lq\lq local moves\rq\rq, the ordinary
Metropolis-Hastings or Gibbs update on every chain; and \lq\lq
global moves\rq\rq\ that include: i) selection of the chains to
swap, based on some probabilistic measures of distance between them;
ii) crossover operator, i.e. partial swap of the current state
between different chains; iii) exchange operator, full state swap
between chains. Both local and global moves are important although
global moves are crucial because they allow the algorithm to jump
from one local mode to another. At the end of the update of
$\bm{\gamma }$, $\tau$ is then sampled using (\ref{T24bis}).

The implementation of EMC that we propose in this paper includes
several novel aspects: the use of a wide range of moves including
two new ones, a local move, based on the Fast Scan
Metropolis-Hastings sampler, particularly suitable when $p$ is large
and a bold global move that exploits the pattern of correlation of
the predictors. Moreover, we developed an efficient scheme for
tuning the temperature placement that capitalises the effective
interchange between the chains. Another new feature is to use a
Metropolis-within-Gibbs with adaptive proposal for updating $\tau$,
as the full conditional (\ref{T24bis}) is not available in closed
form.

\subsection{EMC sampler for $\bm{\gamma }$}
\label{EMC sampler} In what follows, we will only sketch the
rationale behind all the moves that we found useful to implement and
discuss further the benefits of the new specific moves in Section
\ref{Real data examples}. For the \lq\lq large $p$, small $n$\rq\rq\
paradigm and complex predictor spaces, we believe that using a wide
portfolio of moves is needed and offers better guarantee of mixing.

From a notational point of view, we will use the double indexing
$\gamma _{l,j}$, $l=1,\ldots ,L$ and $j=1,\ldots ,p$ to denote the
$j$th latent binary indicator in the $l$th chain. Moreover we
indicate by $\gamma _{l}=\left( \gamma _{l,1},\ldots ,\gamma
_{l,p}\right) ^{T}$ the vector of binary indicators that
characterise the state of the $l$th chain of the population
$\bm{\gamma }=\left( \gamma _{1},\ldots , \gamma _{L}\right) $.

\subsubsection*{Local moves and Fast Scan Metropolis Hastings sampler}
Given $\tau$, we first tried the simple MC$^{3}$ idea of Madigan and
York (1995), also used by Brown \emph{et al.} (1998) where
add/delete and swap moves are used to update the latent binary
vector $\gamma _{l}$. For an add/delete move, one of the $p$
variables is selected at random and if the latent binary value is
$0$ the proposed new value is $1$ or \emph{vice versa}. However,
when $p\gg p_{\gamma _{l}}$, where $p_{\gamma_{l}}$ is the size of
the current model for the $l$th chain, the number of sweeps required
to select by chance a binary indicator with a value of $1$ follows a
geometric distribution with probability $p_{\gamma }/p$ which is
much smaller than $1-p_{\gamma }/p$ to select a binary indicator
with a value of $0$. Hence, the algorithm spends most of the time
trying to add rather than delete a variable. Note that this problem
also affects RJ-type algorithms (Dellaportas \emph{et al.}, 2002).
On the other hand, Gibbs sampling (George and McCulloch, G\&McC
hereafter, 1993) is not affected by this issue since the state of
the $l$th chain is updated by sampling from
\begin{equation}
\left[ p\left( \gamma _{l,j}=1\left\vert y,\gamma _{l,j^{-}},\tau
\right. \right) \right] ^{1/t_{l}}\propto \exp \left\{ \left( \log
p\left( y\left\vert \gamma _{l,j}^{\left( 1\right) },\tau \right.
\right) +\log p\left( \gamma _{l,j}=1\left\vert \gamma
_{l,j^{-}}\right. \right) \right) /t_{l}\right\} ,  \label{T34}
\end{equation}
where $\gamma _{l,j^{-}}$ indicates for the $l$th chain all the
variables, but the $j$th, $j=1,\ldots ,p$ and \\ $\gamma
_{l,j}^{\left( 1\right) }=\left( \gamma _{l,1},\ldots ,\gamma
_{l,j-1},\gamma _{l,j}=1,\gamma _{l,j+1},\ldots ,\gamma
_{l,p}\right) ^{T}$. The main problem related to Gibbs sampling is
the large number of models it evaluates if a full Gibbs cycle or any
permutation of the indices is implemented at each sweep. Each model
requires the direct evaluation, or at least the update, of the time
consuming quantity $S\left( \gamma \right) $, equation (\ref{T6}) or
(\ref{T13}), making practically impossible to rely solely on the
Gibbs sampler when $p$ is very large. However, as sharply noticed by
Kohn \emph{et al.} (2001), it is wasteful to evaluate all the $p$
updates in a cycle because if $p_{\gamma_{l}}$ is much smaller than
$p$ and given $\gamma _{l,j}=0$, it is likely that the sampled value
of $\gamma _{l,j}$ is again $0$.

When $p$ is large, we thus consider instead of the standard MC$^{3}$
add/delete, swap moves, a novel Fast Scan Metropolis-Hastings
scheme, FSMH hereafter, specialised for EMC/PT. It is
computationally less demanding than a full Gibbs sampling on all
$\gamma_{l,j}$ and do not suffer from the problem highlighted before
for MC$^{3}$ and RJ-type algorithms when $p\gg p_{\gamma _{l}}$. The
idea behind the FSMH move is to use an additional
acceptance/rejection step (which is very fast to evaluate) to choose
the number of indices where to perform the Gibbs-like step: the
novelty of our FSMH sampler is that the additional probability used
in the acceptance/rejection step is based not only on the current
chain model size $p_{\gamma _{l}}$, but also on the temperature
$t_l$ attached to the $l$th chain. Therefore the aim is to save
computational time in the large $p$ set-up when multiple chains are
simulated in parallel and finding an alternative scheme to a full
Gibbs sampler. To save computational time our strategy is to
evaluate the time consuming marginal likelihood (\ref{T5}) in no
more than approximately $\left\lfloor
\tilde{\theta}_{l,\bullet}^{(1)}\left( 1/t_{l}\right) \left(
p-p_{\gamma }\right) +\tilde{\theta}_{l,\bullet}^{(0)}\left(
1/t_{l}\right) p_{\gamma }\right\rfloor $ times per cycle in the
$l$th chain (assuming convergence is reached), where
$\tilde{\theta}_{l,\bullet}^{(1)}\left( 1/t_{l}\right) $ is the
probability to select a variable to be added in the
acceptance/rejection step which depends on the current model size
$p_{\gamma _{l}}$ and the temperature $t_l$ and similarly for
$\tilde{\theta}_{l,\bullet}^{(0)}\left( 1/t_{l}\right) $
($\left\lfloor \cdot \right\rfloor $ indicates the integer part).
Since for chains attached to lower temperatures
$\tilde{\theta}_{l,\bullet}^{(0)}\left( 1/t_{l}\right)\gg
\tilde{\theta}_{l,\bullet}^{(1)}\left( 1/t_{l}\right)$, the
algorithm proposes to update \emph{almost all} binary indicators
with value $1$, while it selects at random a group of approximately
$\left\lfloor \tilde{\theta}_{l,\bullet}^{(1)}\left( 1/t_{l}\right)
\left( p-p_{\gamma }\right) \right\rfloor $ binary indicators with
value 0 to be updated. At higher temperatures since
$\tilde{\theta}_{l,\bullet}^{(0)}$ and
$\tilde{\theta}_{l,\bullet}^{(1)}$ become more similar, the number
of models evaluated in a cycle increases because much more binary
indicators with value $0$ are updated. Full details of the FSMH
scheme is given in the Appendix \ref{FSMH scheme}, while evaluation
of them and comparison with MC$^{3}$ embedded in EMC are presented
in Sections \ref{Real data examples} and \ref{Simulation study}

\subsubsection*{Global move: crossover operator}
\label{Crossover operator} The first step of this move consists of
selecting the pair of chains $\left( l,r\right) $ to be operated on.
We firstly compute a probability equal to the weight of the \lq\lq
Boltzmann probability\rq\rq, $p_{t}\left( \gamma _{l}\left\vert \tau
\right. \right) =\exp \left\{ f\left( \gamma _{l}\left\vert \tau
\right. \right) /t\right\} /F_{t}$, where $f\left( \gamma
_{l}\left\vert \tau \right. \right) =\log p\left( \gamma
_{l}\left\vert y,\tau \right. \right) +\log p\left( \gamma
_{l}\right) $ is the log transformation of the full conditional
(\ref{T23bis}) assuming $t_{l}=1$ $\forall l$, $l=1,\ldots ,L$, and
$F_{t}=\sum_{l=1}^{L}\exp \left\{ f\left( \gamma _{l}\left\vert \tau
\right. \right) /t\right\} $ for some specific temperature $t$, and
then rank all the chains according to this. We use normalised
Boltzmann weights to increase the chance that the two selected
chains will give rise, after the crossover, to a new configuration
of the population with higher posterior probability. We refer to
this first step as \lq\lq selection operator\rq\rq.

Suppose that two new latent binary vectors are then generated from
the selected chains according to some crossover operator described
below. The new proposed population of chains \\ $\bm{\gamma
}^{\prime }=\left( \gamma _{1},\ldots ,\gamma _{l}^{\prime},\ldots
,\gamma _{r}^{\prime },\ldots ,\gamma_{L}\right) $ is accepted with
probability
\begin{equation}
\alpha \left( \bm{\gamma }\rightarrow \bm{\gamma }^{\prime }\right)
=\min \left\{ 1,\frac{\exp \left\{ f\left( \gamma _{l}^{\prime
}\left\vert \tau \right. \right) /t_{l}+f\left( \gamma _{r}^{\prime
}\left\vert \tau \right. \right) /t_{r}\right\} }{\exp \left\{
f\left( \gamma _{l}\left\vert \tau \right. \right) /t_{l}+f\left(
\gamma _{r}\left\vert \tau \right. \right) /t_{r}\right\}
}\frac{Q_{t}\left( \bm{\gamma }^{\prime }\rightarrow \bm{\gamma
}\left\vert \tau \right. \right) }{Q_{t}\left( \bm{\gamma
}\rightarrow \bm{\gamma }^{\prime }\left\vert \tau \right. \right)
}\right\}, \label{T29}
\end{equation}
where $Q_{t}\left( \bm{\gamma }\rightarrow \bm{\gamma }^{\prime
}\left\vert \tau \right. \right) $ is the proposal probability, see
Liang and Wong (2000).

In the following we will assume that four different crossover
operators are selected at random at every EMC sweep: $1$-point
crossover, uniform crossover, adaptive crossover (Liang and Wong,
2000) and a novel block crossover. Of these four moves, the uniform
crossover which \lq\lq shuffles\rq\rq\ the binary indicators along
all the selected chains is expected to have a low acceptance, but to
be able to genuinely traverse regions of low posterior probability.
The block crossover essentially tries to swap a group of variables
that are highly correlated and can be seen as a multi-points
crossover whose crossover points are not random but defined from the
correlation structure of the covariates. In practice the block
crossover is defined as follows: one variable is selected at random
with probability $1/p$, then the pairwise correlation $\rho \left(
X_{j},X_{j^{\prime }}\right) $ between the $j$th selected predictor
and each of the remaining covariates, $j^{\prime }=1,\ldots ,p$,
$j^{\prime }\neq j$, is calculated. We then retain for the block
crossover all the covariates with positive (negative) pairwise
correlation with $X_{j}$ such that $\left\vert \rho \left(
X_{j},X_{j^{\prime }}\right) \right\vert \geq \rho _{0}$. The
threshold $\rho _{0}$ is chosen with consideration to the specific
problem, but we fixed it at $0.25$. Evaluation of block crossover
and comparisons with other crossover operators are presented on a
real data example in Section \ref{Real data examples}.

\subsubsection*{Global move: exchange operator}
The exchange operator can be seen as an extreme case of crossover
operator, where the first proposed chain receives the whole second
chain state $\gamma _{l}^{\prime }=\gamma _{r}$, and \emph{vice
versa}. In order to achieve a good acceptance rate, the exchange
operator is usually applied on adjacent chains in the temperature
ladder, which limits its capacity for mixing. To obtain better
mixing, we implemented two different approaches: the first one is
based on Jasra \emph{et al.} (2007) and the related idea of delayed
rejection (Green and Mira, 2001); the second, a bolder \lq\lq
all-exchange\rq\rq\ move, is based on a precalculation of all the
$L\left( L-1\right)/2$ exchange acceptance rates between all chains
pairs (Calvo, 2005). Full relevant details are presented in Appendix
\ref{Exchange}. Both of these bold moves perform well in the real
data applications, see Section \ref{Real data examples}, and
simulated examples, see Section \ref{Simulation study}, thus
contributing to the efficiency of the algorithm.

\subsubsection*{Temperature placement}
As noted by Goswami and Liu (2007), the placement of the temperature
ladder is the most important ingredient in population based MCMC
methods. We propose a procedure for the temperature placement which
has the advantage of simplicity while preserving good accuracy.
First of all, we fix the size $L$ of the population. In doing this,
we are guided by several considerations: the complexity of the
problem, i.e. $E\left( p_{\gamma }\right) $, the size of the data
and computational limits. We have experimented and we recommend to
fix $L\geq 3$. Even though some of the simulated examples had
$p_{\gamma }\simeq 20$ (Section \ref{Simulation study}), we found
that $L=5$ was sufficient to obtain good results. In our real data
examples (Section \ref{Real data examples}), we used $L=4$ guided by
some prior knowledge on $E\left( p_{\gamma }\right) $. Secondly, we
fix at an initial stage, a temperature ladder according to a
geometric scale such that $t_{l+1}/t_{l}=b$, $b>1$, $l=1,\ldots ,L$
with $b$ relatively large, for instance $b=4$. To subsequently tune
the temperature ladder, we then adopt a strategy based on monitoring
only the acceptance rate of the delayed rejection exchange operator
towards a target of $0.5$. Details of the implementation are left in
Appendix \ref{Temperature}

\subsection{Adaptive Metropolis-within-Gibbs for $\tau$}
\label{AMHWG sampler} Various strategies can be used to avoid having
to sample from the posterior distribution of the variable selection
coefficient $\tau$. The easiest way is to integrate it out through a
Laplace approximation (Tierney and Kadane, 1986) or using a
numerical integration such as quadrature on an infinite interval. We
do not pursue these strategies and the reasons can be summarised as
follows. Integrating out $\tau$ in the population implicitly assumes
that every chain has its own value of the variable selection
coefficient $\tau_{l}$ (and of the latent binary vector
$\gamma_{l}$). In this set-up, two unpleasant situations can arise:
firstly, if a Laplace approximation is applied, \emph{equilibrium}
in the product space is difficult to reach because the posterior
distribution of $\gamma _{l} $ depends, through the marginal
likelihood obtained using the Laplace approximation, on the
\emph{chain specific value} of the posterior mode for $\tau_{l}$,
$\hat{\tau}_{\gamma _{l}}$ (details in Appendix
\ref{Laplace_approx}). Since the strength of $X_{\gamma _{l}}$ to
predict the response is weakened for chains attached to high
temperatures, it turns out that for these chains,
$\hat{\tau}_{\gamma _{l}}$ is likely to be close to zero. When the
variable selection coefficient is very small, the marginal
likelihood dependence on $X_{\gamma _{l}}$ decreases even further,
see  for instance (\ref{T6}), and chains attached to high
temperatures will experience a very unstable behaviour, making the
convergence in the product space hard to reach. In addition, if an
automatic tuning of temperature ladder is applied, chains will
increasingly be placed at a closer distance in the temperature
ladder to balance the low acceptance rate of the global moves,
negating the purpose of EMC.

In this paper the convergence is reached instead in the product
space $\prod\nolimits_{l=1}^{L}\left[ p\left( \gamma _{l}\left\vert
y,\tau \right. \right) \right] ^{1/t_{l}}p\left( \tau \right) $,
i.e. the whole population is conditioned on a value of $\tau $
\emph{common to all chains}. This strategy will alleviate the
problems highlighted before allowing for faster convergence and
better mixing among the chains. The procedure just described comes
with an extra cost, i.e. sampling the value of $\tau$. However, this
step is inexpensive in relation to the cost required to sample
$\gamma_{l}$, $l=1,\ldots ,L$. There are several strategies that can
be used to sample $\tau$ from (\ref{T24bis}). We found useful to
apply the idea of adaptive Metropolis-within-Gibbs described in
Roberts and Rosenthal (2008). Conditions for the asymptotic
convergence and ergodicity are guaranteed as we enforce the
\emph{diminishing adaptive condition}, i.e. the transition kernel
stabilises as the number of sweeps goes to infinity and the
\emph{bounded convergence condition}, i.e. the convergence time of
the kernel is bounded in probability. In our set-up using an
adaptive proposal to sample $\tau$ has several benefits; amongst
others it avoids the known problems faced by the Gibbs sampler when
the prior is proper, but relatively flat (Natarajan and McCulloch,
1998) as can happen for Z-S priors when $n$ is large or for the
independent case considered by Bae and Mallick (2004). Moreover,
given an upper limit on the number of sweeps, the adaptation
guarantees a better exploration of the tails of $p\left( \tau
\left\vert y\right. \right) $ than with a fixed proposal. For
details of the implementation and discussion of conditions for
convergence, see Appendix \ref{Adaptive_cond}.

\subsection{ESS algorithm} \label{Algorithm}
In the following, we refer to our proposed algorithm, Evolutionary
Stochastic Search as ESS. If $g$-priors are chosen the algorithm is
denoted as ESS$g$, while we use ESS$i$ if independent priors are
selected (the same notation is used when $\tau$ is fixed or given a
prior distribution). Without loss of generality, we assume that the
response vector and the design matrix have both been centred and, in
the case of independent priors, that the design matrix is also
rescaled. Based on the two full conditionals (\ref{T23bis}) and
(\ref{T24bis}) and the local and global moves introduced earlier,
our ESS algorithm can be summarised as follows.
\begin{itemize}[leftmargin=*, itemsep=-0.5em]
\item Given $\tau$, sample the population's states $\bm{\gamma }$
  from the two steps:
\begin{enumerate}[leftmargin=*, itemsep=-0.5em]
\item[(i)] With probability $0.5$ perform local move and with probability
  $0.5$ apply at random one of the four crossover operators:
  $1$-point, uniform, block and adaptive crossover. If local move is
  selected, use FSMH sampling
  scheme independently for each chain (see Appendix \ref{FSMH scheme}). Moreover
  every $100$ sweeps apply on the first chain a complete
  scan by a Gibbs sampler.
\item[(ii)] Perform the delayed rejection exchange operator or the
  all-exchange operator with equal probability. During the
  burn-in, only select the delayed rejection exchange operator.
\end{enumerate}
\item When $\tau$ is not fixed but has a prior $p\left( \tau \right) $,
  given the latent binary configuration $\bm{\gamma }=\left( \gamma
  _{1},\ldots ,\gamma _{L}\right)$, sample $\tau$ from an
  adaptive Metropolis-within-Gibbs sampling (Section \ref{AMHWG sampler}).
\end{itemize}

From a computational point of view, we used the same fast form for
updating $S\left( \gamma \right) $ as Brown \emph{et al.} (1998),
based on the QR-decomposition. Besides its numerical benefits, QR-
decomposition can deal with the case $p_{\gamma }\geq n$. This
avoids having to restrict the search to models with $p_{\gamma }<n$,
and helps mixing during the burn-in phase.

\section{Performance of ESS} \label{Performance_ESS}
\subsection{Real data examples} \label{Real data examples}
The first real data example is an application of linear regression
to investigate genetic regulation. To discover the genetic causes of
variation in the expression (i.e. transcription) of genes, gene
expression data are treated as a quantitative phenotype while
genotype data (SNPs) are used as predictors, a type of analysis
known as expression Quantitative Trait Loci (eQTL).

\begin{figure}
\subfigure[]{\includegraphics[totalheight=0.25\textheight]{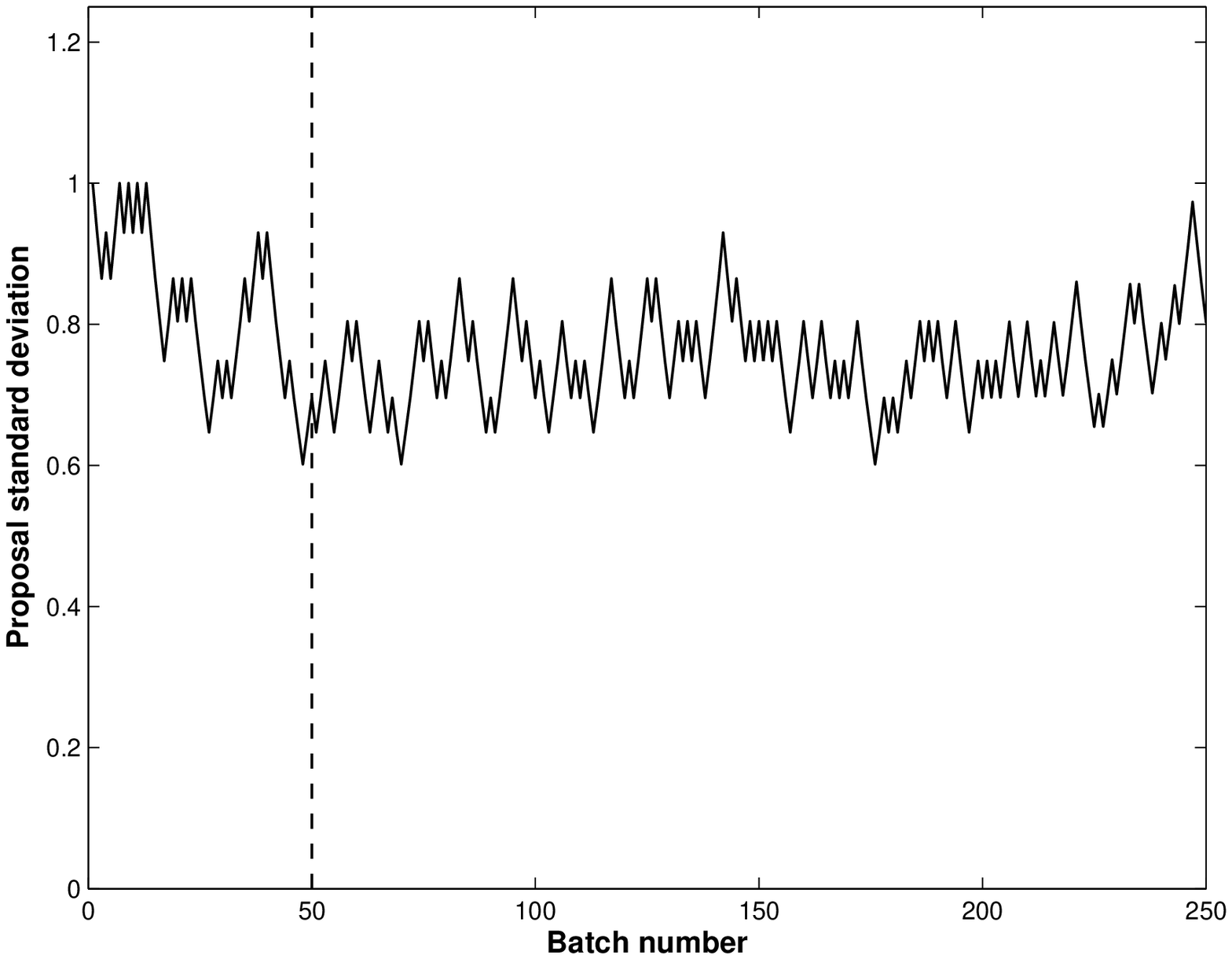}}
\subfigure[]{\includegraphics[totalheight=0.25\textheight]{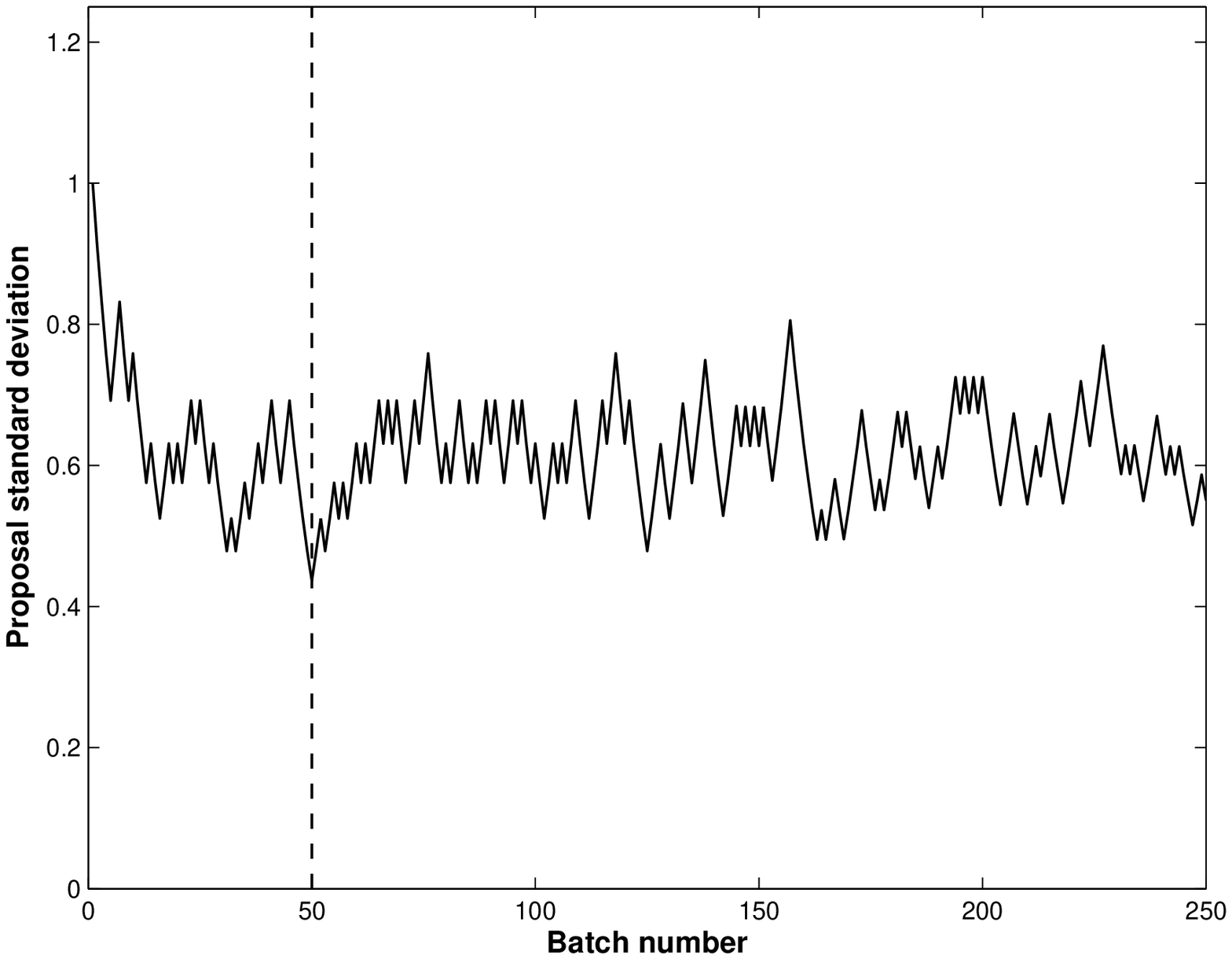}}
\caption{Trace plot of the proposal's standard deviation for $\tau$
for the two real data examples analysed using ESS$g$ with Z-S prior.
Vertical dashed lines indicate the end of the burn-in.}
\label{Fig_S1}
\end{figure}

Here we focus on the ability of ESS to find a parsimonious set of
predictors in an animal data set (Hubner \emph{et al.}, 2005), where
the number of observations, $n=29$, is small with respect to the
number of covariates $p=1,421$. This situation, where $n\ll p$, is
quite common in animal experiments since environmental sources of
variation are controlled as well as the biological diversity of the
sample. For illustration, we report the analysis of one gene
expression response, where we apply ESS$g$ with and without the
hyperprior on $\tau$, see Table \ref{Table_T1}-- eQTL. In the former
case, thanks to the adaptive proposal, the Markov chain for $\tau$
mixes very well reaching an overall acceptance rate which is close
to the target value $0.44$. Convergence issue is not a problem since
the trace of the proposal's standard deviation stabilises quickly
and well inside the bounded conditions, see Figure \ref{Fig_S1}.

\begin{figure}
\begin{center}
\subfigure[]{\includegraphics[totalheight=0.25\textheight]{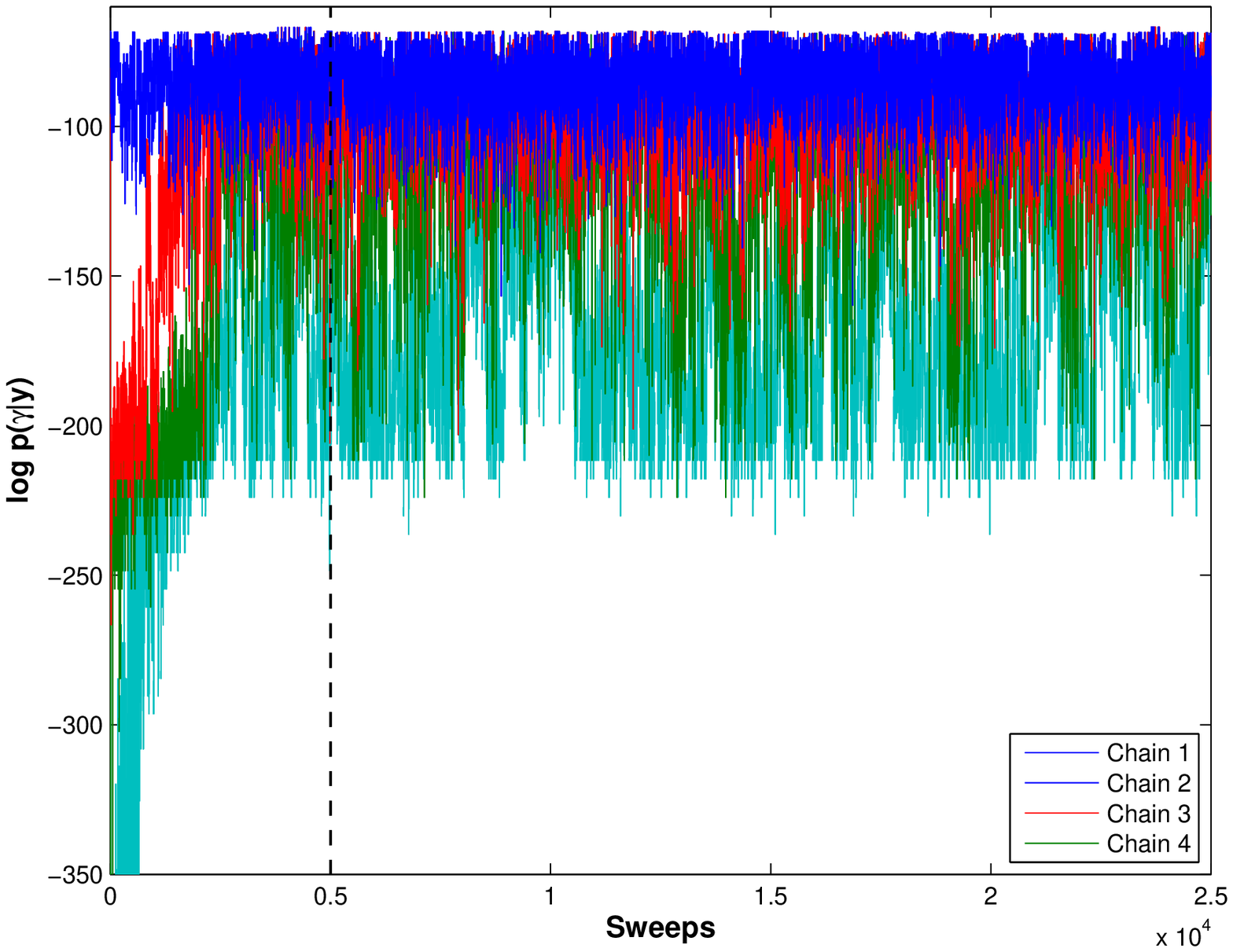}}
\subfigure[]{\includegraphics[totalheight=0.25\textheight]{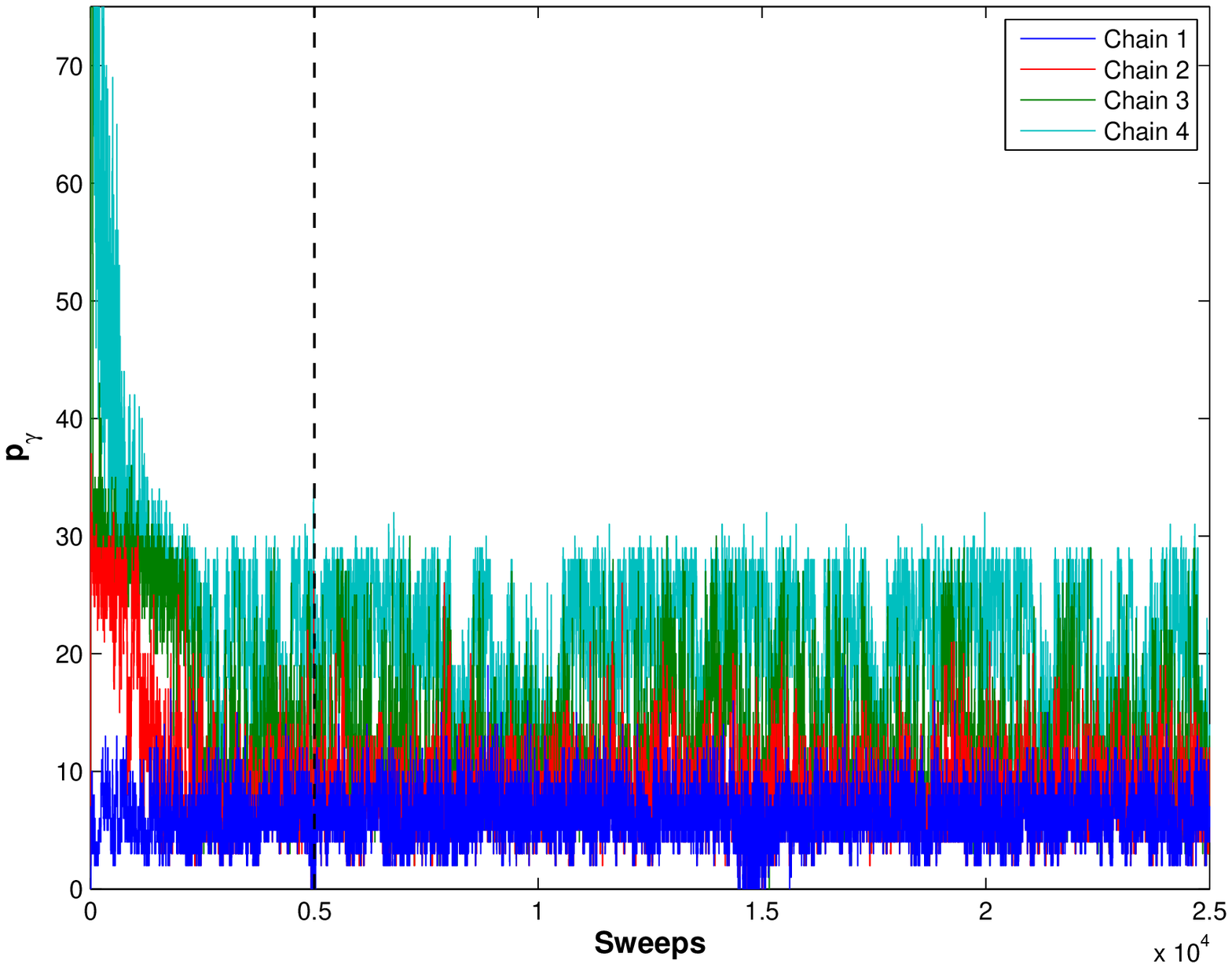}}\\
\subfigure[]{\includegraphics[totalheight=0.25\textheight]{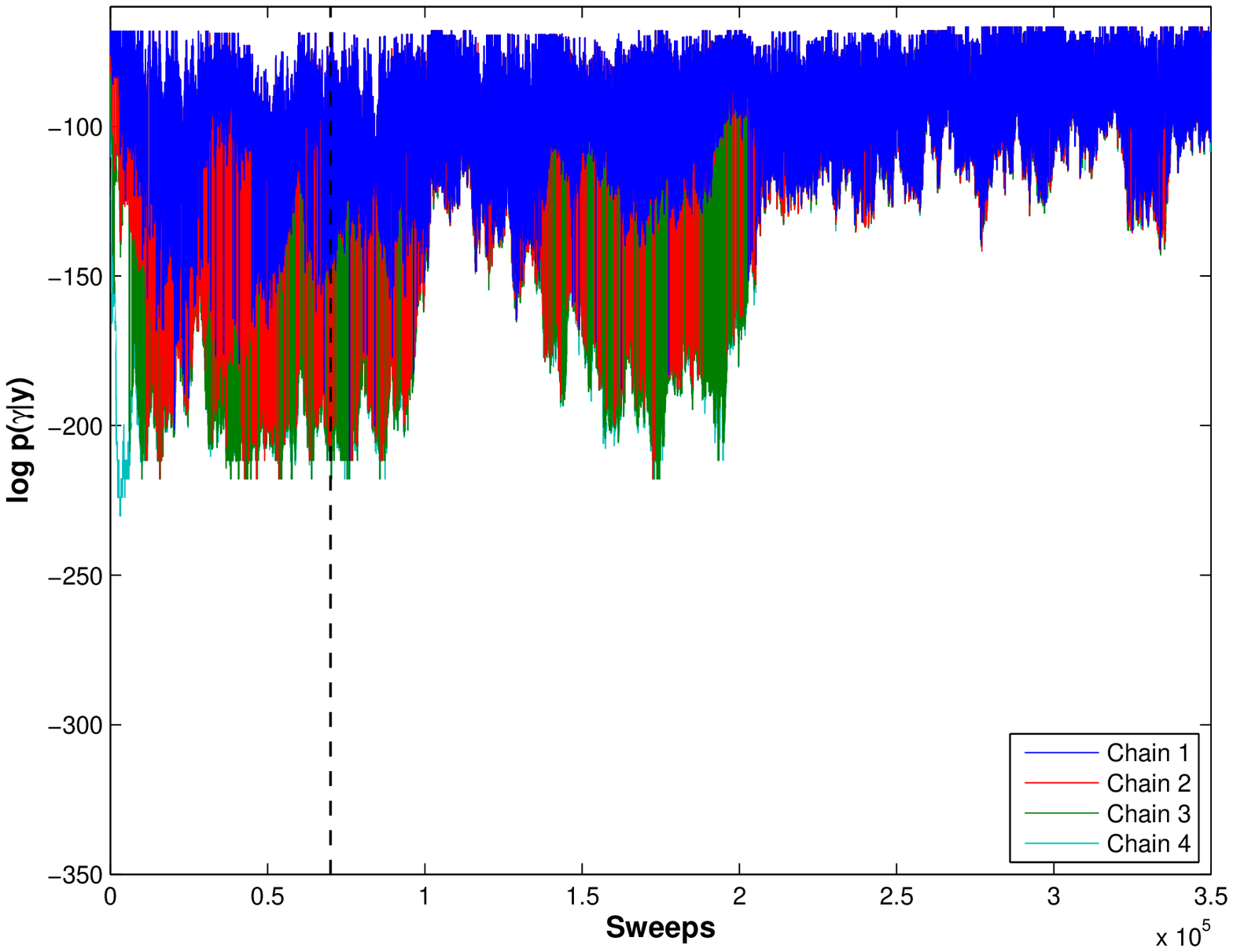}}
\subfigure[]{\includegraphics[totalheight=0.25\textheight]{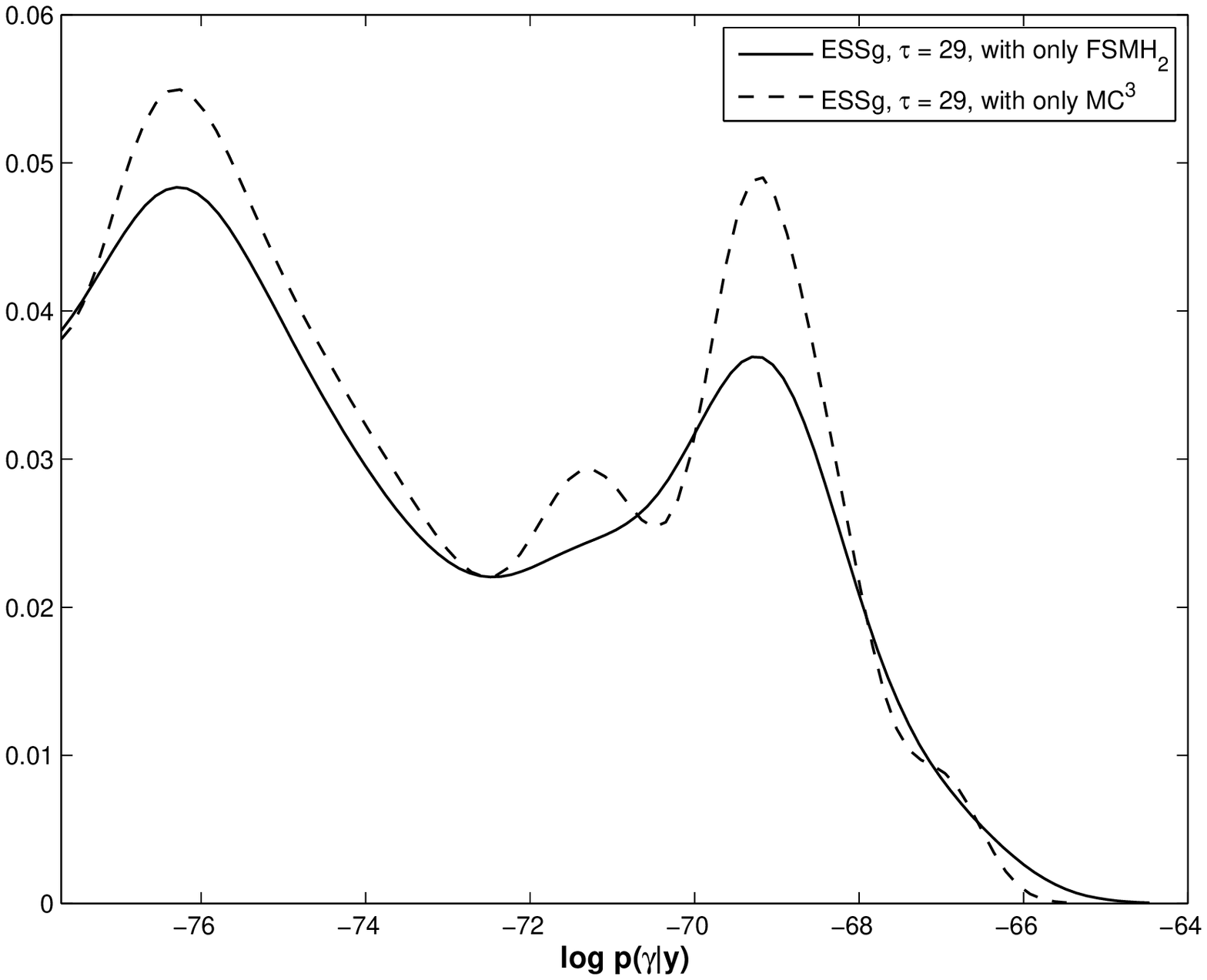}}
\caption{Top panels: (a) trace plot of the log posterior
probability, $\log p\left( \gamma \left\vert y\right. \right) $, and
(b) model size, $p_{\gamma } $, across sweeps for the first real
data example, eQTL analysis, using ESS$g$ with $\tau=29$ and FSMH as
local move. Vertical dashed lines indicate the end of the burn-in.
Bottom panels: (c) trace plot of the log posterior probability when
MC$_{3}$ is used as a local move; (d) kernel densities of $\log
p\left( \gamma \left\vert y\right. \right) $ for the retained chain
in the 25 replicates of the analysis when only FSMH and only
MC$_{3}$ are used as a local move respectively. Plot restricted to
regions of high posterior probability.} \label{Fig_T1}
\end{center}
\end{figure}

In both cases a good mixing among the $L=4$ chains is obtained
(Figure \ref{Fig_T1}, top panels, ESS$g$ with $\tau=29$). Although
in the case depicted in Figure \ref{Fig_T1} with fixed $\tau$, the
convergence is reached in the product space
$\prod\nolimits_{l=1}^{L}\left[ p\left( \gamma _{l}\left\vert
y\right. \right) \right] ^{1/t_{l}}$, by visual inspection we see
that each chain \emph{marginally} reaches its \emph{equilibrium}
with respect to the others; moreover, thanks to the automatic tuning
of the temperature placement during the burn-in, the distributions
of the chains log posterior probabilities overlap nicely, allowing
effective exchange of information between the chains. Table
\ref{Table_T1}--eQTL, confirms that the automatic temperature
selection works well (with and without the hyperprior on $\tau$)
reaching an acceptance rate for the monitored exchange (delayed
rejection) operator close to the selected target of $0.50$. The
all-exchange operator shows a higher acceptance rate, while, in
contrast to Jasra \emph{et al.} (2007), the overall crossover
acceptance rate is reasonable high: in our experience the good
performance of the crossover operator is both related to the
selection operator (Section \ref{EMC sampler}) and the new block
crossover which shows an acceptance rate far higher than the others.
Finally the computational time on the same desktop computer (see
details in Appendix \ref{Performance_comparison}) is rather similar
with or without the hyperprior $\tau$, $28$ and $30$ minutes
respectively for $25,000$ sweeps with $5,000$ as burn-in.

\begin{figure}
\subfigure[]{\includegraphics[totalheight=0.25\textheight]{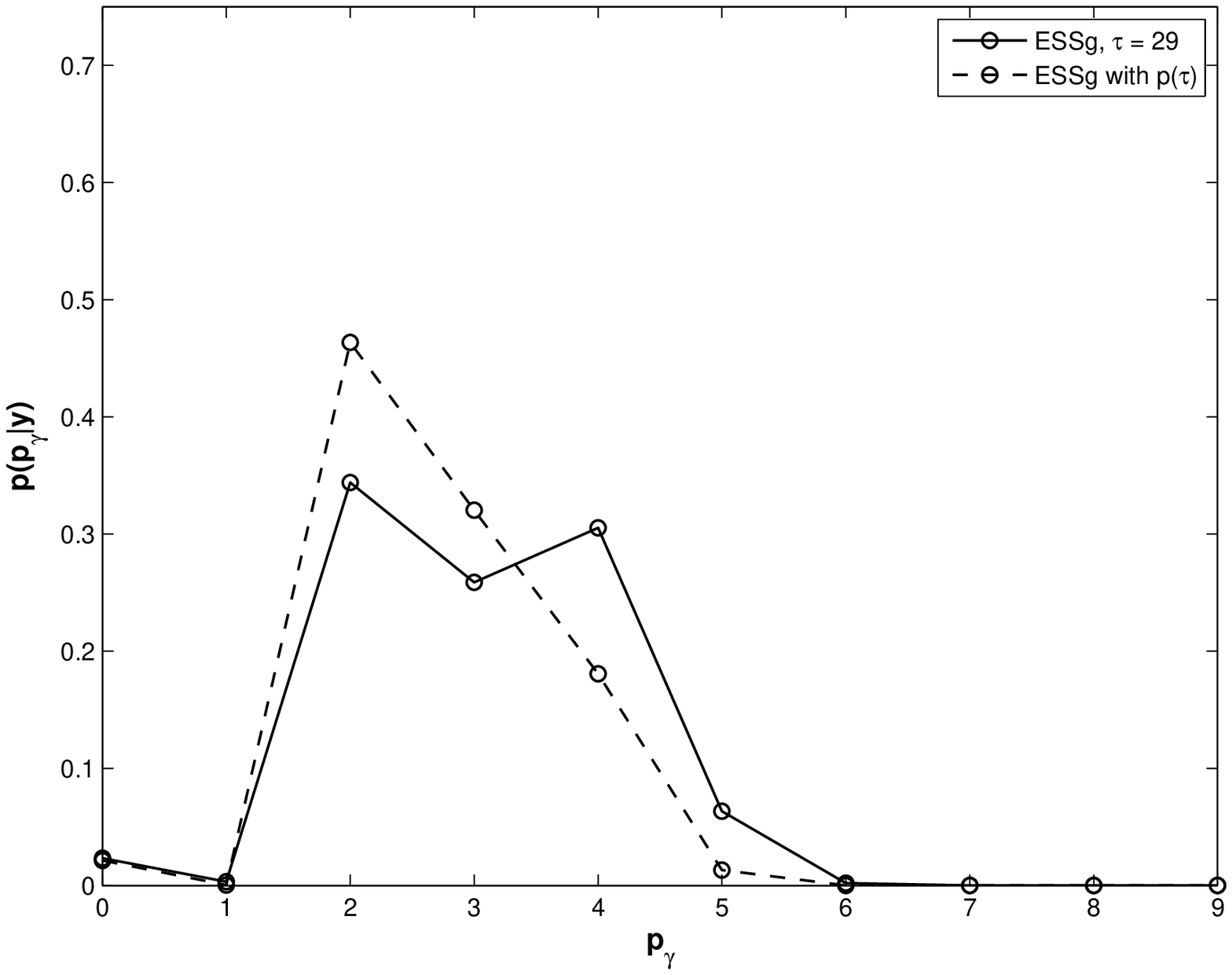}}
\subfigure[]{\includegraphics[totalheight=0.25\textheight]{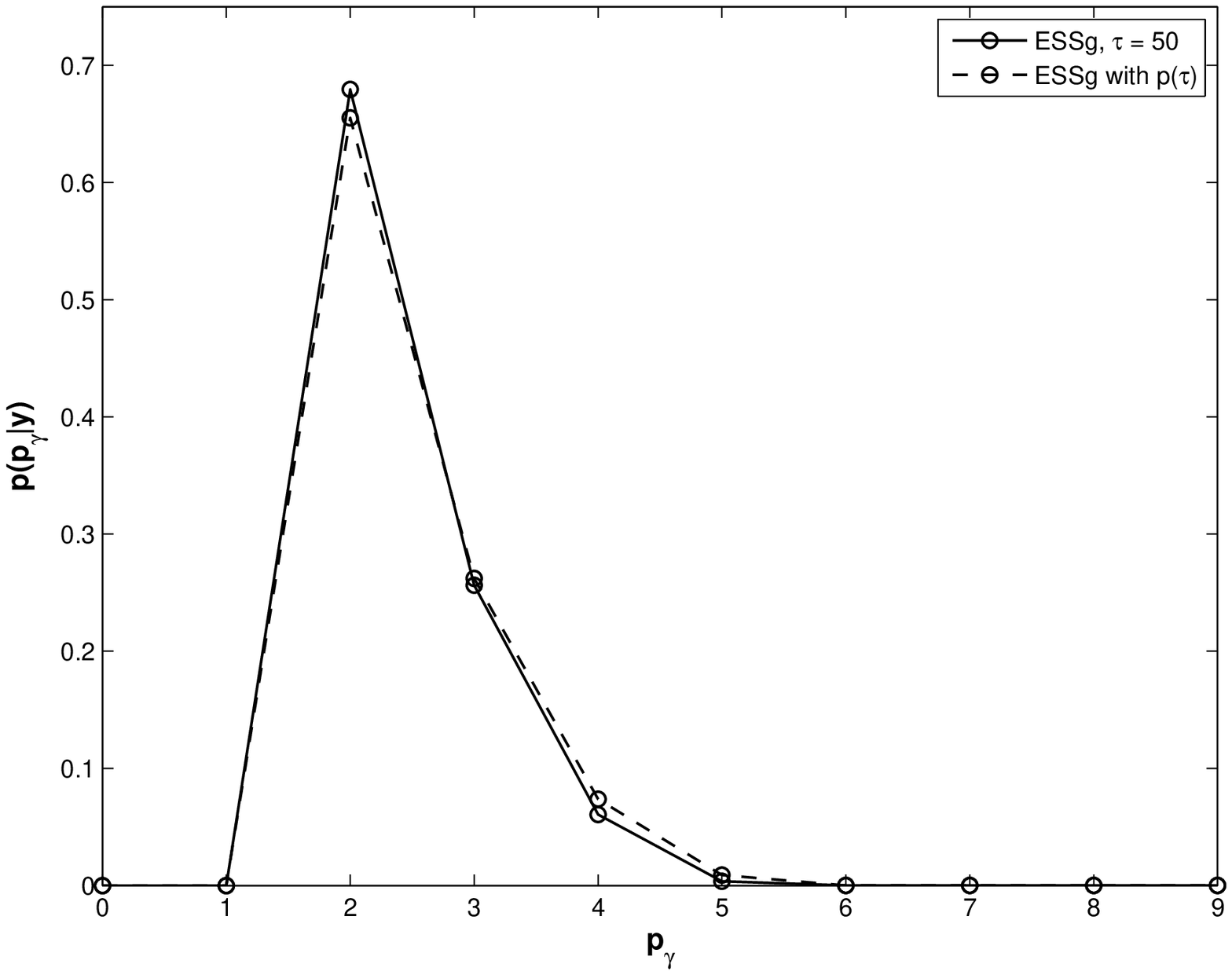}}
\caption{(a) Posterior model size for the first real data example,
eQTL analysis: black solid line for ESS$g$ with $\tau$ fixed at $29$
and black dashed line for ESS$g$ with Z-S prior. (b) Posterior model
size for mQTL analysis, second real data example, using ESS$g$ with
fixed and random $\tau$.} \label{Fig_T2}
\end{figure}

The main difference among the two implementations of ESS$g$ is
related to the posterior model size: when $\tau$ is fixed at $\tau
=29$ (Unit Information Prior, Fern\'{a}ndez \emph{et al.}, 2001),
there is more uncertainty and support for larger models, see Figure
\ref{Fig_T2} (a). In both cases we fix $E\left(p_{\gamma }\right) =
4$ and $V\left( p_{\gamma }\right)=2$, following prior biological
knowledge on the genetic regulation. The posterior mean of the
variable selection coefficient is a little smaller than the Unit
Information Prior, with ESS$g$ coupled with the Z-S prior favouring
smaller models than when $\tau$ is set equal to $29$. The best model
visited (and the corresponding $R_{\gamma }^{2}=1-S(\gamma
)/y^{T}y$) is the same for both version of ESS$g$, while, when a
hyperprior on $\tau$ is implemented, the \lq\lq stability
index\rq\rq\ which indicates how much the algorithm persists on the
first chain top $1,000$ (not unique) visited models ranked by the
posterior probability (Appendix \ref{Performance_comparison}), shows
a higher stability, see Table \ref{Table_T1}-- eQTL. In this case,
having a data-driven level of shrinkage helps the search algorithm
to better discriminate among competing models.

Our second example is related to the application of model (\ref{T1})
in another genomics example: $10,000$ SNPs, selected genome-wide
from a candidate gene study, are used to predict the variation of
Mass Spectography metabolomics data in a small human population, an
example of a so-called mQTL experiment. A suitable dimension
reduction of the data is performed to divide the spectra in regions
or bins and $\log_{10}$-transformation is applied in order to
normalise the signal.

We present the key findings related to a particular metabolite bin,
but the same comments can be extended to the analysis of the whole
data set, where we regressed every metabolites bin  \emph{versus}
the genotype data ($n=50$ and $p=10,000$). In this very challenging
case, we still found an efficient mixing of the chains (see Table
\ref{Table_T1}--mQTL). Note that in this case the posterior mean of
$\tau$, $63.577$, is a little larger than the Unit Information
Prior, $\tau=n$, although the influence of the hyperprior is less
important than in the previous real data example, see Figure
\ref{Fig_T2} (b). In both examples, the posterior model size favours
clearly polygenic control with significant support for up to four
genetic control points (Figure \ref{Fig_T2}) highlighting the
advantage of performing multivariate analysis in genomics rather
than the traditional univariate analysis.

As expected in view of the very large number of predictors, in the
mQTL example the computational time is quite large, around $5$ hours
for $20,000$ sweeps after a burn-in of $5,000$, but as shown in
Table \ref{Table_T1} by the \lq\lq stability index\rq\rq\ ($\approx
0$), we believe that the number of iterations chosen exceeds what is
required in order to visit faithfully the model space. For such
large data analysis tasks, parallelisation of the code could provide
big gains of computer time and would be ideally suited to our
multiple chains approach.

\begin{center}
[Table \ref{Table_T1} about here -- Figure \ref{Fig_T1} about here
-- Figure \ref{Fig_T2} about here -- Figure \ref{Fig_S1} about here]
\end{center}

\begin{figure}
\begin{center}
{\includegraphics[totalheight=0.33\textheight]{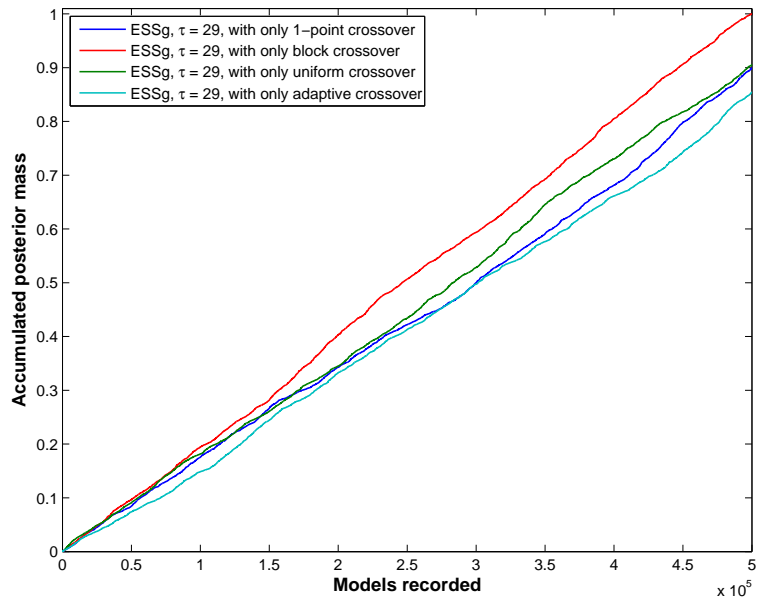}}
\caption{Accumulated posterior mass as a function of the models
recorded. Plot generated using $25$ replicates of the analysis of
the first real data example and normalised by the total mass found
by ESS$g$, $\tau = 29$, with only block crossover move
($\rho_{0}=0.25$). 1-point and uniform crossover accumulate around
$90$\% of the total mass accumulated by ESS$g$ with only block
crossover, while adaptive crossover only $85$\%.} \label{Fig_S2}
\end{center}
\end{figure}

We also evaluate the superiority of our ESS algorithm, and in
particular the FSMH scheme and the block crossover, with respect to
more traditional EMC implementations illustrated for instance in
Liang and Wong (2000). Albeit we believe that using a wide portfolio
of different moves enables any searching algorithm to better explore
complicated model spaces, we reanalysed the first real data example,
eQTL analysis, comparing: (i) ESS$g$ with only FSMH as local move
\emph{vs} ESS with only MC$^{3}$ as local move; (ii) ESS$g$ with
only block crossover \emph{vs} ESS$g$ with only 1-point, only
uniform and only adaptive crossover respectively. To avoid
dependency of the results on the initialisation of the algorithm, we
replicated the analysis $25$ times. Moreover, to make the comparison
fair, in experiment (i) we run the two versions of ESS$g$ for a
different number of sweeps ($25,000$ and $350,000$ with $5,000$ and
$70,000$ as burn-in respectively), but matching the number of models
evaluated. Results are presented in Table \ref{Table_S1}. We report
here the main findings:
\begin{enumerate}[leftmargin=*, itemsep=-0.5em]
\item[(i)] over the $25$ runs, ESS$g$ with FSMH reaches
the same top visited model $68$\% (17/25) while ESS$g$ with MC$^{3}$
the same top model only $28$\%, with a fixed $\tau$, and $88$\% and
$40$\% respectively with Z-S prior. This ability is extended to the
top models ranked by the posterior probability, data not shown,
providing indirect evidence that the proposed new move helps the
algorithm to increase its predictive power. The great superiority
when FSMH scheme are implemented can be explained by comparing
subplot (a) and (c) in Figure \ref{Fig_T1}: the exchange of
information between chains for ESS$g$ with MC$^{3}$ as local move
when $p>n$ (and $p\gg p_{\gamma }$) is rather poor, negating the
purpose of EMC. ESS$g$ with MC$^{3}$ has more difficulties to reach
convergence in the product space and, in contrast to ESS$g$ with
FSMH, the retained chain does not easily escape from local modes.
This later point can be seen looking at Figure \ref{Fig_T1} (d)
which magnifies the right hand tail of the kernel density of $\log
p\left( \gamma \left\vert y\right. \right) $ for the recorded chain,
pulling together the $25$ runs: interestingly ESS$g$ with FSMH is
less \lq\lq bumpy\rq\rq, showing a better ability to escape from
local modes and to explore more efficiently the right tail.
\item[(ii)] Regarding the second comparison when $\tau$ is fixed, ESS$g$
with only block crossover beats constantly the other crossover
operators, with $80$\% \emph{vs} about $60$\%, in terms of best
model visited (Table \ref{Table_S1}) and models with higher
posterior probability (data not shown), has higher acceptance rate
(Table \ref{Table_S2}), showing also a great capacity to accumulate
posterior mass as illustrated in Figure \ref{Fig_S2}. The specific
benefit of the block crossover is less pronounced when a prior on
$\tau$ is specified, but we have already noticed that in this case
having a hyperprior on $\tau$ greatly improves the efficiency of the
search.
\end{enumerate}

\begin{center}
[Table \ref{Table_S1} about here -- Table \ref{Table_S2} about here
-- Figure \ref{Fig_S2} about here]
\end{center}

\subsection{Simulation study} \label{Simulation study}
We briefly report on a comprehensive study of the performance of ESS
in a variety of simulated examples as well as a comparison with SSS.
To make comparison with SSS fair, we use ESS${i}$, the version of
our algorithm which assumes independent priors, $\Sigma _{\gamma
}=\tau I_{p_{\gamma }}$,with $\tau$ fixed at $1$. Details of the
simulated examples (6 set-ups) and how we conducted the simulation
experiment (25 replication of each set-up) are given in Appendix
\ref{Performance_appendix}. The rationale behind the construction of
the examples was to benchmark our algorithm against both $n>p$ and
$p> n$ cases, to use as building blocks intricate correlation
structures that had been used in previous comparisons by G\&McC
(1993, 1997) and N\&G (2004), as well as a realistic correlation
structure derived from genetic data, and to include elements of
model uncertainty in some of the examples by using a range of values
of regression coefficients.

In our example we observe an effective exchange of information
between the chains (reported in Table \ref{Table_S3}) which shows
good overall acceptance rates for the collection of moves that we
have implemented. The dimension of the problem does not seem to
affect the acceptance rates in Table \ref{Table_S3}, remarkably
since values of $p$ range from $60$ to $1,000$ between the examples.
We also studied specifically the performance of the global moves
(Table \ref{Table_S4}) to scrutinise our temperature tuning and
confirmed the good performance of ESS$i$ with good frequencies of
swapping (not far from the case where adjacent chains are selected
to swap at random with equal probability) and good measures of
overlap between chains.

All the examples were run in parallel with ESS${i}$ and SSS 2.0
(Hans \emph{et al.}, 2007) for the same number of sweeps (22,000)
and matching hyperparameters on the model size. Comparison were made
with respect to the marginal probability of inclusion as well as the
ability to reach models with high posterior probability and to
persist in this region. For a detailed discussion of all comparison,
see Appendix \ref{Performance_comparison}.

Overall the covariates with non-zero effects have high marginal
posterior probability of inclusion for ESS$i$ in all the examples,
see Figure \ref{Fig_S4}. There is good agreement between the two
algorithms in general, with additional evidence on some examples
(Figure \ref{Fig_S4} (c) and (d)) that ESS$i$ is able to explore
more fully the model space and in particular to find small effects,
leading to a posterior model size that is close to the true one.
Measures of goodness of fit and stability, Table \ref{Table_S5}, are
in good agreement between ESS$i$ and SSS. The comparison highlight
that a key feature of SSS, its ability to move quickly towards the
right model and to persist on it, is accompanied by a drawback in
having difficulty to explore far apart models with competing
explanatory power, in contrast to ESS$i$ (contaminated example
set-up). Altogether ESS$i$ shows a small improvement of $R_{\gamma
}^{2}$, related to its ability to pick up some of the small effects
that are missed by SSS. Finally ESS$i$ shows a remarkable
superiority in terms of computational time, especially when the
simulated (and estimated) $p_{\gamma }$ is large. Altogether our
comparisons show that we have designed a fully Bayesian MCMC-EMC
sampler which is competitive with the effective search provided by
SSS$i$.

In the same spirit of the real data example analysis, we also
evaluate the superiority of the FSMH scheme with respect to more
traditional EMC implementations, i.e when a MC$^{3}$ local move is
selected. While both versions of the search algorithm visit almost
the same top models ranked by the posterior probability, ESS
persists more on the top models.

\begin{center}
[Table \ref{Table_S3} about here -- Table \ref{Table_S4} about here
-- Table \ref{Table_S5} about here \\
Figure \ref{Fig_S3} about here -- Figure \ref{Fig_S4} about here]
\end{center}

\section{Discussion} \label{Discussion}
The key idea in constructing an effective MCMC sampler for $\gamma$
and $\tau$ is to add an extra parameter, the temperature, that
weakens the likelihood contribution and enables escaping from local
modes. Running parallel chains at different temperature is, on the
other hand, expensive and the added computational cost has to be
balanced against the gains arising from the various \lq\lq
exchanges\rq\rq\ between the chains. This is why we focussed on
developing a good strategy for selecting the pairs of chains, using
both marginal and joint information between the chains, attempting
bold and more conservative exchanges. Combining this with an
automatic choice of the temperature ladder during burn-in is one of
the key element of our ESS algorithm. Using PT in this way has the
potential to be effective in a wide range of situations where the
posterior space is multimodal.

To tackle the case where $p$ is large with respect to $p_{\gamma}$,
the second important element in our algorithm is the use of a
Metropolised Gibbs sampling-like step performed on a subset of
indices in the local updating of the latent binary vector, rather
than an MC$^{3}$ or RJ-like updating move. The new Fast Scan
Metropolis Hastings sampler that we propose to perform these local
moves achieves an effective compromise between full Gibbs sampling
that is not feasible at every sweep when $p$ is large and vanilla
add/delete moves. Comparison of FSMH \emph{vs} MC$^{3}$ scheme on a
real data example and simulation study shows the superiority of our
new local move.

When a model with a prior on the variable selection coefficient
$\tau$ is preferred, the updating of $\tau$ itself present no
particular difficulties and is computationally inexpensive.
Moreover, using an adaptive sampler makes the algorithm self
contained without any time consuming tuning of the proposal
variance. This latter strategy works perfectly well both in the
$g$-prior and independent prior case as illustrated in Sections
\ref{Real data examples} and \ref{Simulation study}. Our current
implementation does not make use of the output of the heated chains
for posterior inference. Whether gains in variance reduction could
be achieved in the spirit of Gramacy \emph{et al.} (2007) is an area
for further exploration, which is beyond the scope of the present
work.

Our approach has been applied so far to linear regression with
univariate response $y$. An interesting generalisation is that of a
multidimensional $n \times q$ response $Y$ and the identification of
regressors that jointly predict the $Y$ (Brown \emph{et al.}, 1998).
Much of our set-up and algorithm carries through without
difficulties and we have already implemented our algorithm in this
framework in a challenging case study in genomics with
multidimensional outcomes.

\vspace{-0.30cm}
\section*{Acknowledgements} The authors are thankful to Norbert Hubner
and Timothy Aitman for providing the data of the eQTL example,
Gareth Roberts and Jeffrey Rosenthal for helpful discussions about
adaptation and Michail Papathomas for his detailed comments. Sylvia
Richardson acknowledges support from the MRC grant GO.600609.

\vspace{-0.30cm}
\section*{Appendix}
\begin{appendix}
\renewcommand{\theequation}{A.\arabic{equation}}
\setcounter{equation}{0}

\section{Technical details of EMC implementation} \label{Tecnical
point} In this Section we will describe some technical details
omitted from the paper and related to the sampling schemes we used
for the population of binary latent vectors $\bm{\gamma}$ and the
selection coefficient $\tau$.

\subsection{EMC sampler for $\bm{\gamma }$}
\subsubsection*{Local move: FSMH scheme} \label{FSMH scheme} Let $\gamma _{l,j}$, $l=1,\ldots ,L$ and
$j=1,\ldots ,p$ to denote the $j$th latent binary indicator in the
$l$th chain. As in Kohn \emph{et al.} (2001), let $\gamma
_{l,j}^{\left( 1\right) }=\left( \gamma _{l,1},\ldots ,\gamma
_{l,j-1},\gamma
_{l,j}=1,\gamma _{l,j+1},\ldots ,\gamma _{l,p}\right) ^{T}$ and \\
$\gamma _{l,j}^{\left( 0\right) }=\left( \gamma _{l,1},\ldots
,\gamma _{l,j-1},\gamma _{l,j}=0,\gamma _{l,j+1},\ldots ,\gamma
_{l,p}\right) ^{T}$. Furthermore let $L_{l,j}^{\left(
1\right)}\propto p\left( y\left\vert \gamma _{l,j}^{\left( 1\right)
},\tau\right. \right) $ and $L_{l,j}^{\left( 0\right)}\propto
p\left( y\left\vert \gamma _{l,j}^{\left( 0\right) },\tau\right.
\right) $ and finally $\theta _{l,j}^{\left( 1\right)}=p\left(
\gamma _{l,j}=1\left\vert \gamma _{l,j^{-}}\right. \right) $ and
$\theta _{l,j}^{\left( 0\right)}=1-\theta _{l,j}^{\left( 1\right)}$.
From (\ref{T7}) it is easy to prove that
\begin{equation}
\theta _{l,j}^{\left( 1\right)}=p\left( \gamma _{l,j}=1\left\vert
\gamma _{l,j^{-}}\right. \right) =\frac{p_{\gamma _{l}}+a_{\omega
}-1}{p+a_{\omega }+b_{\omega }-1}, \label{AL1}
\end{equation}
where $p_{\gamma _{l}}$ is the current model size for the $l$th
chain. Using the above equation, for $\gamma _{l,j}=1$ the
normalised version of (\ref{T34}) can be written as
\begin{equation}
\left[ p\left( \gamma _{l,j}=1\left\vert y,\gamma _{l,j^{-}},\tau
\right. \right) \right] ^{1/t_{l}}=\frac{\left. \theta
_{l,j}^{\left( 1\right) }\right. ^{1/t_{l}}\left. L_{l,j}^{\left(
1\right) }\right. ^{1/t_{l}}}{S\left( 1/t_{l}\right) }, \label{AL2}
\end{equation}
where $S\left( 1/t_{l}\right) =\left. \theta _{l,j}^{\left( 1\right)
}\right.^{1/t_{l}}\left. L_{l,j}^{\left( 1\right) }\right.
^{1/t_{l}}+\left. \theta _{l,j}^{\left( 0\right) }\right.
^{1/t_{l}}\left. L_{l,j}^{\left( 0\right) }\right. ^{1/t_{l}}$ with
$\left[ p\left( \gamma _{l,j}=1\left\vert y,\gamma _{l,j^{-}},\tau
\right. \right) \right] ^{1/t_{l}}$ defined similarly. Hence if
$\left. \theta _{l,j}^{\left( 1\right) }\right. ^{1/t_{l}}$ is very
small, then $\left[ p\left( \gamma _{l,j}=1\left\vert y,\gamma
_{l,j^{-}},\tau \right. \right) \right] ^{1/t_{l}}$ is small as
well. Therefore for the Gibbs sampler with a beta-binomial prior on
the model space, the posterior probability of $\gamma _{l,j}=1$
depends crucially on $\left.\theta _{l,j}^{\left( 1\right)
}\right.^{1/t_{l}}$.

In the following we derive a Fast Scan Metropolis-Hastings scheme
specialised for Evolutionary Monte Carlo or parallel tempering. We
define $Q\left( 1\rightarrow 0\right) =Q\left( \gamma _{l,j}^{\left(
1\right) }\rightarrow \gamma _{l,j}^{\left( 0\right) }\right) $ as
the proposal probability to go from $1$ to $0$ and $Q\left(
0\rightarrow 1\right) $ the proposal probability to go from $0$ to
$1$ for the $j$th variable and $l$th chain. Moreover using the
notation introduced before, the Metropolis-within-Gibbs version of
(\ref{T34}) to go from $0$ to $1$ in the EMC local move is
\begin{equation}
\alpha _{l}^{\text{MwG}}\left( 0\rightarrow 1\right) =\min \left\{
1,\frac{ \left. \theta _{l,j}^{\left( 1\right) }\right.
^{1/t_{l}}\left. L_{l,j}^{\left( 1\right) }\right.
^{1/t_{l}}}{\left. \theta _{l,j}^{\left( 0\right) }\right.
^{1/t_{l}}\left. L_{l,j}^{\left( 0\right) }\right.
^{1/t_{l}}}\frac{Q\left( 1\rightarrow 0\right) }{Q\left(
0\rightarrow 1\right) }\right\} \label{AL3}
\end{equation}
with a similar expression for $\alpha _{l}^{\text{MwG}}\left(
1\rightarrow 0\right) $. The proof of the Propositions are omitted
since they are easy to check. We first introduce the following
Proposition which is useful for the calculation of the acceptance
probability in the FSMH scheme.

\begin{Prop}  \label{Prop1}
The following three conditions are equivalent: a) $\left.
L_{l,j}^{\left( 0\right) }\right. ^{1/t_{l}} \left.
\left/L_{l,j}^{\left( 1\right) }\right. ^{1/t_{l}}\geq 1\right.$;
\\ b) $\left. L_{l,j}^{\left( 1\right) }\right. ^{1/t_{l}}\left/
\tilde{S}\left( 1/t_{l}\right) \geq1\right.$; c)$\left.
L_{l,j}^{\left( 0\right) }\right. ^{1/t_{l}}\left/ \tilde{S}\left(
1/t_{l}\right)<1\right.$, where $\tilde{S}\left( 1/t_{l}\right)
=S\left( 1/t_{l}\right) \left/ \left( \left. \theta _{l,j}^{\left(
1\right) }\right. ^ {1/t_{l}}+\left. \theta _{l,j}^{\left( 0\right)
}\right. ^{1/t_{l}}\right) \right. $ is the convex combination of
the marginal likelihood $\left. L_{l,j}^{\left( 1\right) }\right.
^{1/t_{l}}$ and $\left. L_{l,j}^{\left( 0\right) }\right.
^{1/t_{l}}$ with weights $\tilde{\theta}_{l,j}^{\left(
1\right)}\left( 1/t_{l}\right) =\left. \theta _{l,j}^{\left(
1\right) }\right. ^{1/t_{l}}\left/ \left( \left. \theta
_{l,j}^{\left( 1\right) }\right. ^{1/t_{l}}+\left. \theta
_{l,j}^{\left( 0\right) }\right. ^{1/t_{l}}\right) \right.$ and
$\tilde{\theta}_{l,j}^{\left( 0\right)}\left( 1/t_{l}\right)
=1-\tilde{\theta} _{l,j}^{\left( 1\right)}\left( 1/t_{l}\right) $.
\end{Prop}

The FSMH scheme can be seen as a random scan Metropolis-within-Gibbs
algorithm where the number of evaluations is linked to the
prior/current model size and the temperature attached to the chain.
The computation requirement for the additional acceptance/rejection
step is very modest since the normalised tempered version of
(\ref{AL1}) is used.

\begin{Prop}  \label{Prop3}
Let $l=1,\ldots ,L$, $j=1,\ldots ,p$ (or any permutation of them),
$Q^{\text{FSMH}}\left( 0\rightarrow 1\right)
=\tilde{\theta}_{l,j}^{\left( 1\right)}\left( 1/t_{l}\right)$ and
$Q^{\text{FSMH}}\left( 1\rightarrow 0\right)
=\tilde{\theta}_{l,j}^{\left( 0\right)}\left( 1/t_{l}\right) $ with
$\tilde{\theta}_{l,j}^{\left( 0\right)}\left( 1/t_{l}\right)
=1-\tilde{\theta} _{l,j}^{\left( 1\right)}\left( 1/t_{l}\right) $.
The acceptance probabilities are
\begin{equation}
\alpha _{l}^{\text{FSMH}}\left( 0\rightarrow 1\right) =\left\{
\begin{tabular}{ll}
$1$ & $\text{if }\left. L_{l,j}^{\left( 1\right) }\right.
^{1/t_{l}}\left/\left. L_{l,j}^{\left( 0\right) }\right.
^{1/t_{l}}\right.\geq 1$
\\ $\left. L_{l,j}^{\left( 1\right) }\right. ^{1/t_{l}}\left/\left.
L_{l,j}^{\left( 0\right) }\right. ^{1/t_{l}}\right.$ & $\text{if }
\left. L_{l,j}^{\left( 1\right) }\right. ^{1/t_{l}}\left/\left.
L_{l,j}^{\left( 0\right) }\right. ^{1/t_{l}}\right.<1$
\end{tabular}
\right. \label{AL6}
\end{equation}
\begin{equation}
\alpha _{l}^{\text{FSMH}}\left( 1\rightarrow 0\right) =\left\{
\begin{tabular}{ll}
$1$ & $\text{if }\left. L_{l,j}^{\left( 0\right) }\right.
^{1/t_{l}}\left/\left.
L_{l,j}^{\left( 1\right) }\right. ^{1/t_{l}}\right.\geq 1$ \\
$\left. L_{l,j}^{\left( 0\right) }\right. ^{1/t_{l}}\left/\left.
L_{l,j}^{\left( 1\right) }\right. ^{1/t_{l}}\right.$ & $\text{if }
\left. L_{l,j}^{\left( 0\right) }\right. ^{1/t_{l}}\left/\left.
L_{l,j}^{\left( 1\right) }\right. ^{1/t_{l}}<1\right.$
\end{tabular}
\right. \label{AL7}
\end{equation}
\end{Prop}
The above sampling scheme works as follows. Given the $l$th chain,
if $\gamma _{lj}=0$ (and similarly for $\gamma _{lj}=1$), it
proposes the new value from a Bernoulli distribution with
probability $\tilde{\theta}_{l,j}^{\left( 1\right)}\left(
1/t_{l}\right) $: if the proposed value is different from the
current one, it evaluates (\ref{AL6}) (and similarly
\ref{AL7})otherwise it selects a new covariate.

Finally it can be proved that the Gibbs sampler is more efficient
than the FSMH scheme, i.e. for a fixed number of iterations, Gibbs
sampling MCMC standard error is lower than for FSMH sampler. However
the Gibbs sampler is computationally more expensive so that, if $p$
is very large, as described in Kohn \emph{et al.} (2001), FSMH
scheme becomes more efficient per floating point operation.

\subsubsection*{Global move: exchange operator} \label{Exchange}
The exchange operator can be seen as an extreme case of crossover
operator, where the first proposed chain receives the whole second
chain state $\gamma _{l}^{\prime }=\gamma _{r}$, and the second
proposed chain receives the whole first state chain $\gamma
_{r}^{\prime }=\gamma _{l}$, respectively.

In order to achieve a good acceptance rate, the exchange operator is
usually applied on adjacent chains in the temperature ladder, which
limits its capacity for mixing. To obtain better mixing, we
implemented two different approaches: the first one is based on
Jasra \emph{et al.} (2007) and the related idea of delayed rejection
(Green and Mira, 2001); the second one on Gibbs distribution over
all possible chains pairs (Calvo, 2005).
\begin{enumerate}[leftmargin=*, itemsep=0em]
\item The delayed rejection exchange operator tries first to swap
the state of the chains that are usually far apart in the
temperature ladder, but, once the proposed move has been rejected,
it performs a more traditional (uniform) adjacent pair selection,
increasing the overall mixing between chains on one hand without
drastically reducing the acceptance rate on the other. However its
flexibility comes at some extra computational costs and in
particular the additional evaluation of the pseudo move necessary to
maintain detailed balance (Green and Mira, 2001). Details are
reported below.

Suppose two chains are selected at random, $l$ and $r$ with $l\neq
r$, in order to swap their binary latent vector. Then, given that
$\gamma _{l}^{\prime }=\gamma _{r}$, $\gamma _{r}^{\prime
}=\gamma_{l}$ and $Q_{t}\left( \bm{\gamma }\rightarrow \bm{\gamma
}^{\prime }\right) =Q_{t}\left( \bm{\gamma }^{\prime }\rightarrow
\bm{\gamma }\right) $, (13) reduces to
\[
\alpha _{1}\left( \bm{\gamma }\rightarrow \bm{\gamma }^{\prime
}\right) =\min \left\{ 1,\frac{\exp \left\{ f\left( \gamma
_{r}\left\vert \tau\right. \right) /t_{l}+f\left( \gamma
_{l}\left\vert \tau\right. \right) /t_{r}\right\} }{\exp \left\{
f\left( \gamma _{l}\left\vert \tau\right. \right) /t_{l}+f\left(
\gamma _{r}\left\vert \tau\right. \right) /t_{r}\right\} }\right\}.
\]
Since the two chains are selected at random, the above acceptance
probability decreases exponentially with the difference $\left\vert
1/t_{l}-1/t_{r}\right\vert $ and therefore most of the proposed
moves are rejected. If rejected, a delayed rejection-type move is
applied between two random adjacent chains, with $l$ the first one
and $s$, $\left\vert l-s\right\vert =1$, the second one, giving rise
to the new acceptance probability
\[
\alpha _{2}\left( \bm{\gamma }\rightarrow \bm{\gamma }^{\prime
\prime }\right) =\min \left\{ 1,\frac{\exp \left\{ f\left( \gamma
_{s}\left\vert \tau\right. \right) /t_{l}+f\left( \gamma
_{l}\left\vert \tau\right. \right) /t_{s}\right\} }{\exp \left\{
f\left( \gamma _{l}\left\vert \tau\right. \right) /t_{l}+f\left(
\gamma _{s}\left\vert \tau\right. \right) /t_{s}\right\}
}\frac{1-\alpha _{1}\left( \bm{\gamma }^{\prime \prime }\rightarrow
\bm{\gamma }^{\ast }\right) }{1-\alpha _{1}\left( \bm{ \gamma
}\rightarrow \bm{\gamma }^{\prime }\right) }\right\},
\]
where the pseudo move $\bm{\gamma }^{\ast }$ is necessary in order
to maintain the detailed balance condition (Green and Mira, 2001).

\item Alternatively, we attempt a bolder \lq\lq
  all-exchange\rq\rq\ operator. Swapping the state of two chains that
  are far apart in the temperature ladder speeds up the convergence of
  the simulation since it replaces several adjacent swaps with a
  single move. However, this move can be seen as a rare event whose
  acceptance probability is low and unknown. Since the full set of
  possible exchange moves is finite and discrete, it is easy and
  computationally inexpensive to calculate all the $L\left( L-1\right)
  /2$ exchange acceptance rates between all chains pairs, inclusive
  the rare ones, $\tilde{p}_{l,r}=\exp \left\{ \left( f\left( \gamma
  _{r}\left\vert \tau \right. \right) -f\left( \gamma _{l}\left\vert
  \tau \right. \right) \right) \left( 1/t_{l}-1/t_{r}\right) \right\}
  $. To maintain detailed balance condition, the possibility not to
  perform any exchange (rejection) must be added with unnormalised
  probability one. Finally the chains whose states are swopped are
  selected at random with probability equal to
\begin{equation}
p_{h}=\frac{\tilde{p}_{h}}{\sum_{h=1}^{1+L\left( L-1\right)
/2}\tilde{p}_{h}}, \label{T32}
\end{equation}
where in (\ref{T32}) each pair $\left( l,r<l\right) $ is denoted by
a single number $h$, $\tilde{p}_{h}=\tilde{p}_{l,r}$, including the
rejection move, $h=1$.
\end{enumerate}

\subsubsection*{Temperature placement} \label{Temperature}
First we select the number $L$ of chains close to the complexity of
the problem, i.e. $E\left( p_{\gamma }\right) $, although the size
of the data and computational limits need to be taken into account.
Secondly, we fix a first stage temperature ladder according to a
geometric scale such that $t_{l+1}/t_{l}=b$, $b>1$, $l=1,\ldots ,L$
with $b$ relatively large, for instance $b=4$. Finally, we adopt a
strategy similar to the one described in Roberts and Rosenthal
(2008), but \emph{restricted to the burn-in stage}, monitoring only
the acceptance rate of the delayed rejection exchange operator.
After the $k$th \lq\lq batch\rq\rq\ of EMC sweeps, to be chosen but
usually set equal to $100$, we update $b_k$, the value of the
constant $b$ up to the $k$th batch, by adding or subtracting an
amount $\delta _{b}$ such that the acceptance rate of the delayed
rejection exchange operator is as close as possible to $0.50$ (Liu,
2001; Jasra \emph{et al.}, 2007), $b_{k+1}=2^{\log _{2}b_{k}\pm
\delta _{b}}$. Specifically the value of $\delta_{b}$ is chosen such
that at the end of the burn-in period the value of $b$ can be 1. To
be precise, we fix the value of $\delta _{b}$ as $\log _{2}\left(
b_{1}\right) /\tilde{K}$, where $b_{1}$ is the first value assigned
to the geometric ratio and $\tilde{K}$ is the total number of
batches in the burn-in period.

\subsection{Adaptive Metropolis-within-Gibbs for $\tau$}
\subsubsection*{Laplace approximation for the conditional marginal
likelihood} \label{Laplace_approx} Under model (1) and prior
specification for $\alpha $, (2) and (3), we provide the Laplace
approximation of $p\left( y\left\vert \gamma ,\tau \right. \right) $
for the $g$-prior case, while the approximation for the independent
case can be derived following the same line of reasoning. For easy
of notation we drop the chain subscript index and we assume that the
observed responses $y$ have been centred with mean $0$, i.e. $\left(
y-\bar{y}_{n}\right) \equiv y$. In the following we will distinguish
the cases in which the posterior mode $\hat{\tau}_{\gamma }$ is a
solution of a cubic or quadratic equation. Conditions on the
existence of the solutions are provided as well as those that
guarantee the positive semidefiniteness of the variance
approximation. Recall that
\begin{eqnarray*}
p\left( y\left\vert \gamma \right. \right)  &=&\int \exp \left\{
\log \left( p\left( y\left\vert \gamma ,\tau\right. \right) p\left(
\tau\right) \right)
\right\} d\tau \\
&\approx &\sqrt{2\pi }\sigma _{\hat{\lambda}}\left( \log p\left(
y\left\vert \gamma ,\hat{\lambda}\right. \right) +\log p\left(
\hat{\lambda}\right) +\log J\left( \hat{\lambda}\right) \right),
\end{eqnarray*}
where $\hat{\lambda}$ is the posterior mode after the transformation
$\lambda =\log \left( \tau\right) $, which is necessary to avoid
problems on the boundary, $\sigma _{\hat{\lambda}}$ is the
approximate squared root of the variance calculated in
$\hat{\lambda}$ and $J\left( \cdot \right) $ is the Jacobian of the
transformation. Details about Laplace approximation can be found in
Tierney and Kadane (1986). Similar derivations when $p\left( \sigma
^{2}\right) \propto \sigma ^{-2}$ are presented in Liang \emph{et
al.} (2008). Finally throughout the presentation we will assume that
$n>p_{\gamma }$ and that $a_{g}$ and $b_{g}$ are fixed small as in
Kohn \emph{et al.} (2001).

\noindent \underline{Cubic equation for Zellner-Siow priors} \\ If
$p\left( \tau\right) = InvGa\left( a_{\tau},b_{\tau}\right) $ the
posterior $\hat{\lambda}$ mode is the only positive root of the
integrand function
\begin{equation*}
I_{\lambda }=\left( 1+e^{\lambda }\right) ^{\left( 2a_{\sigma
}+n-1-p_{\gamma }\right) /2}\left\{ 2b_{\sigma }\left( 1+e^{\lambda
}\right) +y^{T}y\left[ 1+e^{\lambda }\left( 1-R_{\gamma }^{2}\right)
\right] \right\} ^{-\left( 2a_{\sigma }+n-1\right)/2
}\frac{e^{-b_{\tau}/e^{\lambda }}}{\left( e^{\lambda }\right)
^{a_{\tau}+1}}e^{\lambda },
\end{equation*}
where the last factor in the above equation $e^{\lambda }=\left\vert
de^{\lambda }/d\lambda \right\vert $ is the Jacobian of the
transformation. After the calculus of the first derivative of the
log transformation and some algebra manipulations, it can be shown
that $e^{\hat{\lambda}}$ is the solution of the cubic equation
\begin{equation}
e^{3\lambda }+\frac{c_{1}c_{3}-c_{2}c_{4}-\left( c_{3}+c_{4}\right)
a_{\tau}+c_{4}b_{\tau}}{\left( c_{1}-c_{2}-a_{\tau}\right)
c_{4}}e^{2\lambda }+\frac{ -c_{3}a_{\tau}+\left( c_{3}+c_{4}\right)
b_{\tau}}{\left( c_{1}-c_{2}-a_{\tau}\right) c_{4}}e^{\lambda
}+\frac{c_{3}b_{\tau}}{\left( c_{1}-c_{2}-a_{\tau}\right) c_{4}}=0
\label{AL11}
\end{equation}
and that
\begin{eqnarray}
\sigma _{\hat{\lambda}}^{2} &=&\left. -\frac{1}{\left( \log p\left(
y\left\vert \gamma ,\lambda \right. \right) +\log p\left( \lambda
\right)
\right) ^{\prime \prime }}\right\vert _{\lambda =\hat{\lambda}}  \notag \\
&=&\left[ -c_{1}\frac{e^{\lambda }}{\left( 1+e^{\lambda }\right)
^{2}}+c_{2} \frac{c_{3}c_{4}e^{\lambda }}{\left(
c_{3}+c_{4}e^{\lambda }\right) ^{2}}+ \frac{b_{\tau}}{e^{\lambda
}}\right] _{\lambda =\hat{\lambda}}^{-1}, \label{AL2_bis}
\end{eqnarray}
where $c_{1}=\left( 2a_{\sigma }+n-1-p_{\gamma }\right) /2$,
$c_{2}=\left( 2a_{\sigma }+n-1\right) /2$, $c_{3}=2b_{\sigma
}+y^{T}y$ and $c_{4}=2b_{\sigma }+y^{T}y\left( 1-R_{\gamma
}^{2}\right) $. Following Liang \emph{et al.} (2008), since
$\lim_{\lambda \rightarrow -\infty }\partial I_{\lambda }/\partial
\lambda >0 $, because $c_{3}b_{\tau}>0$, and $\lim_{\lambda
\rightarrow \infty }\partial I_{\lambda }/\partial \lambda <0 $,
because $\left( c_{1}-c_{2}-a_{\tau}\right) c_{4}<0$, at least one
real positive solution exists. Moreover since $-\left(
c_{3}b_{\tau}\right) /\left( c_{1}-c_{2}-a_{\tau}\right) c_{4}>0$,
the remaining two real solutions should have the same sign
(Abramowitz and Stegun, 1970). A necessary condition for the
existence of just one real positive solution is that the summation
of all the pairs-products of the coefficients is negative
\begin{equation*}
\frac{-c_{3}a_{\tau}+\left( c_{3}+c_{4}\right) b_{\tau}}{\left(
c_{1}-c_{2}-a_{\tau}\right) c_{4}}<0
\end{equation*}
and this happens if $b_{\tau}/a_{\tau}>c_{3}/\left(
c_{3}+c_{4}\right) $. When $R_{\gamma }^{2}\rightarrow 0$ and thus
$c_{3}=c_{4}$, the above condition corresponds to
$b_{\tau}>a_{\tau}/2$ and when $R_{\gamma }^{2}\rightarrow 1$, as
$c_{3}/\left( c_{3}+c_{4}\right) \approx 1$ especially when $y^{T}y$
is large, which might be expected when $n$ becomes large, the
condition is equivalent to $b_{\tau}>a_{\tau}$. Therefore it turns
out that a sufficient condition for the existence of just one real
positive solution in (\ref{AL1}) is $b_{\tau}>a_{\tau}$.

The positive semidefiniteness of the approximate variance can be
proved as follows. First of all it is worth noticing that all the
terms in (\ref{AL2_bis}) are of the same order $O_p\left(
e^{-\lambda }\right) $. Then, when $R_{\gamma }^{2}\rightarrow 0$,
the positive semidefiniteness is always guaranteed, while when
$R_{\gamma }^{2}\rightarrow 1$, provided that $y^{T}y$ is large, the
middle term in (\ref{AL2_bis}) tends to zero and the condition is
fulfilled if $b_{\tau}>c_{1}$.

\noindent \underline{Quadratic equation for Liang \emph{et al.} (2008) prior} \\
If $p\left( \tau\right) \propto \left( 1+\tau\right) ^{-c_{\tau}}$,
with $c_{\tau}>0$, $e^{\hat{\lambda}}$ is only the positive root of
the integrand function
\[
I_{\lambda }=\left( 1+e^{\lambda }\right) ^{\left( 2a_{\sigma
}+n-1-p_{\gamma }-c_{\tau }\right) /2}\left\{ 2b_{\sigma }\left(
1+e^{\lambda }\right) +y^{T}y\left[ 1+e^{\lambda }\left( 1-R_{\gamma
}^{2}\right) \right] \right\} ^{-\left( 2a_{\sigma }+n-1\right)
/2}e^{\lambda }
\]
or, after the first derivative of the log transformation, the
solution of the quadratic equation
\begin{equation}
\left( c_{1}^{\ast }-c_{2}+1\right) c_{4}e^{2\lambda }+\left(
c_{1}^{\ast }c_{3}-c_{2}c_{4}+c_{3}+c_{4}\right) e^{\lambda
}+c_{3}=0  \label{AL3_bis}
\end{equation}
with $c_{1}^{\ast }=\left[ 2a_{\sigma }+n-1-\left( p_{\gamma
}+2c_{\tau}\right) \right] /2$ and $c_{2}$, $c_{3}$ and $c_{4}$
defined as above. The discriminant of the quadratic equation is
$\Delta =\left( c_{1}^{\ast
}c_{3}-c_{2}c_{4}c_{3}+c_{3}+c_{4}\right) ^{2}-4\left( c_{1}^{\ast
}-c_{2}+1\right) c_{4}c_{3}$ which is always greater than zero and
therefore two real roots exist. Since one of them is positive in
order to prove that (\ref{AL3_bis}) admits just one positive
solution, it is necessary to show that
\begin{equation*}
\frac{-\left( c_{1}^{\ast }c_{3}-c_{2}c_{4}+c_{3}+c_{4}\right)
-\Delta ^{1/2} }{2\left( c_{1}^{\ast }-c_{2}+1\right) c_{4}}<0
\end{equation*}
which is true provided that $\left( c_{1}^{\ast }-c_{2}+1\right)
c_{4}c_{3}<0$. Moreover the approximate variance can be written as
\begin{equation}
\sigma _{\hat{\lambda}}^{2}=\left[ -c_{1}^{\ast }\frac{e^{\lambda
}}{\left( 1+e^{\lambda }\right)
^{2}}+c_{2}\frac{c_{3}c_{4}e^{\lambda }}{\left(
c_{3}+c_{4}e^{\lambda }\right) ^{2}}\right] _{\lambda
=\hat{\lambda}}^{-1} \label{AL4_bis}
\end{equation}
which is positive semidefinite when $R_{\gamma }^{2}\rightarrow 0$
if $c_{2}>c_{1}^{\ast }$, which is always verified, while, if
$R_{\gamma }^{2}\rightarrow 1$ and $y^{T}y$ is large, equation
(\ref{AL4_bis}) is not positive unless $p_{\gamma
}+2c_{\tau}>2a_{\sigma }+n-1$.

The explicit solution of the posterior mode is also available
\begin{eqnarray}
\hat{\tau}_{\gamma } &=&\max \left\{ \frac{\left( c_{4}-c_{3}\right)
/\left( c_{1}^{\ast
}-c_{2}\right) }{c_{4}/c_{1}^{\ast }}-1,0\right\}   \notag \\
&=&\max \left\{ \frac{R_{\gamma }^{2}/\left( p_{\gamma
}+2c_{\tau}\right) }{ \left[ 2b_{\sigma }/\left( y^{T}y\right)
+\left( 1-R_{\gamma }^{2}\right) \right] /\left[ 2a_{\sigma
}+n-1-\left( p_{\gamma }+2c_{\tau}\right) \right] } -1,0\right\}
\label{AL13}
\end{eqnarray}
which corresponds to MLE if $c_{\tau}=0$.

\subsubsection*{Diminishing adaptive and bounded conditions}
\label{Adaptive_cond} Since $\tau$ is defined on the real positive
axis we propose the new value of $\tau$ on the logarithm scale. In
particular we use as proposal the normal distribution centred at the
current value of $\log \left( \tau\right) $ in the $g$-prior and
independent prior case. The variance of the proposal distribution is
controlled as illustrated in Roberts and Rosenthal (2008): every
$100$ EMC sweeps, the same value of sweeps used in the temperature
placement, we monitor the acceptance rate of the
Metropolis-within-Gibbs algorithm: if it is lower (higher) than the
optimal acceptance rate, i.e. 0.44, a constant $\delta _{\tau}(k)$
is added (subtracted) to $ls_{k}$, the log standard deviation of the
proposal distribution in the $k$th batch of EMC sweeps. The value of
the constant to be added or subtracted is rather arbitrary, but we
found useful to fix it as $\left\vert ls_{1}-5\right\vert
/\tilde{K}$, where $\tilde{K}$ is the total number of batches in the
burn-in period: during the burn-in the log standard deviation should
be able to reach any values at a distance $\pm 5$ in log scale from
the initial value of $ls_{1}$ usually set equal to zero. The
\emph{diminishing adaptive condition} is obtained imposing $\delta
_{\tau}\left( k\right) =\min \{\left\vert ls_{1}-5\right\vert
/\tilde{K} ,k^{-1/2}\}$, where $k$ is the current number of batches,
including the burn-in. To ensure the \emph{bounded convergence
condition} we follow Roberts and Rosenthal (2008), restricting each
$ls_{k}$ to be inside $\left[ M_{1},M_{2}\right] $ and we fix them
equal to $M_{1}=-10$ and $M_{2}=10$ respectively. In practise these
bounds do not create any restriction since the sequence of the
standard deviations of the proposal distribution stabilises almost
immediately, indicating that the transition kernel converges in a
bounded number of batches, see Figure \ref{Fig_T2}.

\section{Performance of ESS: Simulation study} \label{Performance_appendix}
In this Section we report in details on the performance of ESS in a
variety of simulated examples. Main conclusions are summarised in
the Section \ref{Simulation study}.

Firstly we analyse the simulated examples with ESS${i}$ the version
of our algorithm which assumes independent priors, $\Sigma _{\gamma
}=\tau I_{p_{\gamma }}$, so as to enable comparisons with SSS which
also implements an independent prior. Moreover, in order to make to
comparison with SSS fair, in the simulation study only the first
step of the algorithm described in Section 3.3 is performed, with
$\tau$ fixed at $1$. As in SSS, standardisation of the covariates is
done before running ESS${i}$. We run ESS${i}$ and SSS 2.0 (Hans
\emph{et al.}, 2007) for the same number of sweeps (22,000) and with
matching hyperparameters on the model size.

Secondly, to  discuss the mixing properties of ESS  when a prior
$p\left( \tau \right) $ is defined on $\tau$, we implement both the
$g$-prior and independent prior set-up for a particular simulated
experiment. To be precise in the former case we will use the
Zellner-Siow priors (\ref{T9}), and for the latter we will specify a
proper but diffuse exponential distribution as suggested by Bae and
Mallick (2004).

\subsection{Simulated experiments} \label{Simulated experiments}
We apply ESS with independent priors to an extensive and challenging
range of simulated examples with $\tau$ fixed at $1$: the first
three examples (Ex1-Ex3) consider the case $n>p$ while the remaining
three (Ex4-Ex6) have $p> n$. Moreover in all examples, except the
last one, we simulate the design matrix, creating more and more
intricated correlation structures between the covariates in order to
test the proposed algorithm in different and increasingly more
realistic scenarios. In the last example, we use, as design matrix,
a genetic region spanning $500$-kb from the HapMap project
(Altshuler \emph{et al.}, 2005).

Simulated experiments Ex1-Ex5 share in common the way we build $X$.
In order to create moderate to strong correlation, we found useful
referring to two simulated examples in George and McCulloch, G\&McC
hereafter, (1993) and in G\&McC (1997): throughout we call $X_{1}$
($n\times 60$) and $X_{2}$ $(n\times15)$ the design matrix obtained
from these two examples. In particular the $j$th column of $X_{1}$,
indicated as $X_{\left( 1\right) j}$, is simulated as $X_{\left(
1\right) j}=X_{j}^{\ast }+Z$, where $X_{1}^{\ast },\ldots
,X_{60}^{\ast }$ iid $\sim N_{n}\left( 0,1\right) $ independently
form $Z\sim N_{n}\left( 0,1\right) $, inducing a pairwise
correlation of $0.5$. $X_{2}$ is generated as follows: firstly we
simulated $Z_{1},\ldots ,Z_{15}$ iid $\sim N_{n}\left( 0,1\right) $
and we set $X_{\left( 2\right) j}=Z_{i}+2Z_{j}$ for
$j=1,3,5,8,9,10,12,13,14,15$ only. To induce strong
multicollinearity, we then set $X_{\left( 2\right) 2}=X_{\left(
2\right) 1}+0.15Z_{2}$, $X_{\left( 2\right) 4}=X_{\left( 2\right)
3}+0.15Z_{4}$, $X_{\left( 2\right) 6}=X_{\left( 2\right)
5}+0.15Z_{6}$, $X_{\left( 2\right) 7}=X_{\left( 2\right)
8}+X_{\left( 2\right) 9}-X_{\left( 2\right) 10}+0.15Z_{7}$ and
$X_{\left( 2\right) 11}=X_{\left( 2\right) 14}+X_{\left( 2\right)
15}-X_{\left( 2\right) 12}-X_{\left( 2\right) 13}+0.15Z_{11}$. A
pairwise correlation of about 0.998 between $X_{\left( 2\right) j}$
and $X_{\left( 2\right) j+1}$ for $j=1,3,5$ is introduced and
similarly strong linear relationship is present within the sets
$\left( X_{\left( 2\right) 7},X_{\left( 2\right) 8},X_{\left(
2\right) 9},X_{\left( 2\right) 10}\right) $ and $\left( X_{\left(
2\right) 11},X_{\left( 2\right) 12},X_{\left( 2\right) 13},X_{\left(
2\right) 14},X_{\left( 2\right) 15}\right) $.

Then, as in Nott and Green, N\&G hereafter, (2004) Example 2, more
complex structures are created by placing side by side combinations
of $X_{1}$ and/or $X_{2}$, with different sample size. We will vary
the number of samples $n$ in $X_{1}$  and $X_{2}$ as we construct
our examples. The levels of $\beta $ are taken from the simulation
study of Fern\'{a}ndez \emph{et al.} (2001), while the number of
true effects, $p_\gamma$, with the exception of Ex3, varies from $5$
to $16$. Finally the simulated error variance ranges from $0.05^{2}$
to $2.5^{2}$ in order to vary the level of difficulty for the search
algorithm. Throughout we only list the non-zero $\beta_{\gamma}$ and
assume that $\beta _{\gamma^{-} }=0^{T}$. The six examples can be
summarised as follows:

\begin{itemize}[leftmargin=*, itemsep=-0.5em]
\item [\textbf{Ex1}:] $X=X_{1}$ is a matrix of dimension
  $120\times 60$, where the responses are simulated from (1)
  using $\alpha =0$, $\gamma =\left( 21,37,46,53,54\right) ^{T}$,
  $\beta _{\gamma }=\left( 2.5,0.5,-1,1.5,0.5\right) ^{T}$,
  and $\varepsilon \sim N\left(
  0,2^{2}I_{120}\right) $. In the following we will not refer to the
  intercept $\alpha $ any more since, as described in Section
  3.3 in the paper, we consider $y$ centred and hence there is
  no difference in the results if the intercept is simulated or not.
  This is the simplest of our example, although, as reported in G\&McC
  (1993) the average pairwise correlation is about $0.5$, making it
  already hard to analyse by standard stepwise methods.
\item [\textbf{Ex2}:]This example is taken directly from N\&G
  (2004), Example 2, who first introduce the idea of combining simpler
  \lq\lq building blocks\rq\rq\ to create a new matrix $X$ : in their
  example $X=\left[ X_{2}^{\left( 1\right) }X_{2}^{\left( 2\right)
  }\right] $ is a $300\times 30$ matrix, where $X_{2}^{\left( 1\right)
  }$ and $X_{2}^{\left( 2\right) }$ are of dimension  $300\times 15$
  and have each the same structure as $ X_{2}$. Moreover $\gamma
  =\left( 1,3,5,7,8,11,12,13\right) ^{T}$, $\beta _{\gamma }=\left(
  1.5,1.5,1.5,1.5,-1.5,1.5,1.5,1.5\right) ^{T}$ and $\varepsilon \sim
  N\left( 0,2.5^{2}I_{300}\right) $. We chose this example for two
  reasons: firstly, since the correlation structure in $X_{2}$ is very
  involved, we test the proposed algorithm under strong and
  complicated correlations between the covariates; secondly, since $y$
  is not simulated from the second \lq\lq block\rq\rq, we are
  interested to see if the proposed algorithm does \emph{not} select
  any variable that belongs to the second group.
\item [\textbf{Ex3}:]As in G\&McC (1993), Example 2, $X=X_{1}$,
  is a $120\times 60$ matrix, $\beta =\left( \beta _{1},\ldots ,\beta
  _{60}\right) ^{T}$, $\left( \beta _{1},\ldots ,\beta _{15}\right)
  =\left( 0,\ldots ,0\right) $, $\left( \beta _{16},\ldots ,\beta
  _{30}\right) =\left( 1,\ldots ,1\right) $, $\left( \beta
  _{31},\ldots ,\beta _{45}\right) =\left( 2,\ldots ,2\right) $,
  $\left( \beta _{46},\ldots ,\beta _{60}\right) =\left( 3,\ldots
  ,3\right) $ and $\varepsilon \sim N\left( 0,2^{2}I_{120}\right) $.
  The motivation behind this example is to test the strength of the
  proposed algorithm to select a subset of variables which is large
  with respect to $p$ while preserving the ability \emph{not} to
  choose any of the first $15$ variables.
\item [\textbf{Ex4}:]The design matrix $X$, $120\times 300$, is
  constructed as follows: firstly we create a new $120\times60$ \lq\lq
  building block\rq\rq, $X_{3}$, combining $X_{2}$ and a smaller
  version of $X_{1}$, $X_{1}^{\ast }$, a $120\times 45$ matrix
  simulated as $X_{1}$, such that $X_{3}=\left[ X_{2}X_{1}^{\ast
  }\right] $ (dimension $120\times 60$). Secondly we place side by side five copies of $X_{3}$,
  $X=\left[ X_{3}^{\left( 1\right) }X_{3}^{\left(2\right)
  }X_{3}^{\left( 3\right) }X_{3}^{\left( 4\right) }X_{3}^{\left(
  5\right) }\right] $: the new design matrix alternates blocks of
  covariates of high and complicated correlation, as in G\&McC (1997),
  with regions where the correlation is moderate as in G\&McC (1993).
  We simulate the response selecting $16$ variables from $X$, \\ $\gamma
  =\left(
  1,11,30,45,61,71,90,105,121,131,150,165,181,191,210,225\right) ^{T}$
  such that every pair belongs alternatively to $X_{2}$ or $X_{1}$. We
  simulate $y$ using \\ $\beta _{\gamma }=\left(
  2,-1,1.5,1,0.5,2,-1,1.5,1,0.5,2,-1,-1,1.5,1,0.5\right) ^{T}$ with
  $\varepsilon \sim N\left( 0,2.5^{2}I_{120}\right) $. This example is
  challenging in view of the correlation structure, the number of
  covariates $p > n$ and the different levels of the effects.
\item [\textbf{Ex5}:]This is the most challenging example that we
  simulated and it is based on the idea of  contaminated models. The
  matrix $X$, $200\times 1000$, is $X=\left[ X_{3}^{\left( 1\right)
  }X_{3}^{\left( 2\right) }X_{3}^{\left( 3\right) }X_{1}^{\ast \ast
  }X_{3}^{\left( 4\right) }X_{3}^{\left( 5\right) }X_{3}^{\left(
  6\right) }X_{3}^{\left( 7\right) }X_{3}^{\left( 8\right) } \right]
  $, with $X_{1}^{\ast \ast }$, a $200\times 520$ larger version of
  $X_{1}$. We partitioned the responses such that $y=[y_{1}
  y_{2}]^{T}$: $y_1$ is simulated from \lq\lq model 1\rq\rq\ ($\gamma
  ^{1}=\left( 701,730,745,763,790,805,825,850,865,887\right) $ and
  $\beta _{\gamma }^{1}=\left( 2,-1,1.5,1,0.5,2,-1,1.5,2,-1\right) $)
  while $y_2$ is simulated from \lq\lq model 2\rq\rq\ ($\gamma
  ^{2}=\left( 1,38,63,98,125\right) $ and $\beta _{\gamma }^{2}=\left(
  2,-1,1.5,1,0.5\right) $). Finally, fixing $\varepsilon \sim N\left(
  0,0.05^{2}I_{200}\right) $ and the sample size in the two models
  such that $y_{1}$ and $y_{2}$ are vectors of dimension $1\times 160$
  and $1\times 40$ respectively, $y$ is retained if, given the
  sampling variability, we find $R_{\gamma ^{1}}^{2}\geq 0.6$ and
  $R_{\gamma ^{1}}^{2}/8\leq R_{\gamma ^{2}}^{2}\leq R_{\gamma
  ^{1}}^{2}/10$: in this way we know that \lq\lq model 1\rq\rq\
  accounts for most of the variability of $y$, but without a
  negligible effect for \lq\lq model 2\rq\rq. In this example, we
  measure the ability of the proposed algorithm to recognise the most
  promising model and therefore being robust to contaminations.
  However since ESS can easily jump between local modes we are also
  interested to see if \lq\lq model 2\rq\rq\ is selected.
\item [\textbf{Ex6}:]The last simulated example is based on phased genotype
  data from HapMap project (Altshuler \emph{et al.}, 2005), region
  ENm014, Yoruba population: the data set originally contained 1,218
  SNPs (Single Nucleotide Polymorphism) for 120 chromosomes, but after eliminating redundant
  variables, the design matrix reduced to $120\times 775$. While in the previous
  examples a \lq\lq block structure\rq\rq\ of correlated variables is
  artificially constructed, in this example blocks of linkage
  disequilibrium (LD) derive naturally from genetic forces, with a
  slow decay of the level of pairwise correlation between SNPs.
  Finally we chose $\gamma =\left(
  50,75,140,200,300,400,500,650,700,770\right) ^{T}$ such that the
  effects are visually inside blocks of LD, with their size simulated
  from $\beta _{\gamma }\sim N\left( 0,3^2I_{10}\right) $ with
  $\varepsilon \sim N\left( 0,0.10^{2}I_{120}\right) $. Since the simulated
  effects can range roughly between
  $ \left( -6, 6\right) $, this will allow us to test also the ability
  of ESS$i$ to select small effects.
\end{itemize}

We conclude this Section by reporting how we conducted the
simulation experiment: every example from Ex1 to Ex6 has been
replicated $25$ times and the results presented for example Ex1 to
Ex5 are averaged over the $25$ replicates. For Ex6 the effects size
change so average across replicated is only done for the mixing
properties. ESS$i$  with  $\tau$ =1 was applied to each
example/sample, recording the visited sequence of $\gamma_{1}$ for
$20,000$ sweeps after a burn-in of $2,000$ required for the
automatic tuning of the temperature placement, Section \ref{EMC
sampler} With the exception of Ex2 and Ex3, where we used an
indifferent prior, $p\left( \gamma \right) =\left( 1/2\right) ^{p}$,
we analysed the remaining examples setting $E\left( p_{\gamma
}\right) =5$ with $V\left( p_{\gamma }\right) =E\left( p_{\gamma
}\right) \left( 1-E\left( p_{\gamma }\right) /p\right) $ which
corresponds to a binomial prior over $p_{\gamma }$. In order to
establish the sensitivity of the proposed algorithm to the choice of
$E\left( p_{\gamma }\right) $ we also analysed Ex1 fixing $E\left(
p_{\gamma }\right) =10$ and $20$. Moreover in all the examples we
chose $L=5$ with the starting value of $\bm{\gamma }$ chosen at
random. The remaining two hyperparameters to be fixed, namely
$a_{\sigma }$ and $b_{\sigma }$, are set equal to $a_{\sigma
}=10^{-6}$ and $b_{\sigma }=10^{-3}$ as in Kohn \emph{et al.} (2001)
which corresponds to a relative uninformative prior.

\subsection{Mixing properties of ESS${i}$} \label{Mixing properties}
\begin{figure}
\begin{center}
\subfigure[]{\includegraphics[totalheight=0.25\textheight]{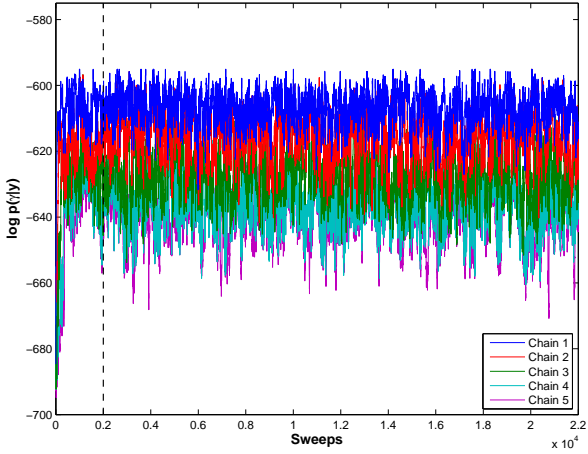}}
\subfigure[]{\includegraphics[totalheight=0.25\textheight]{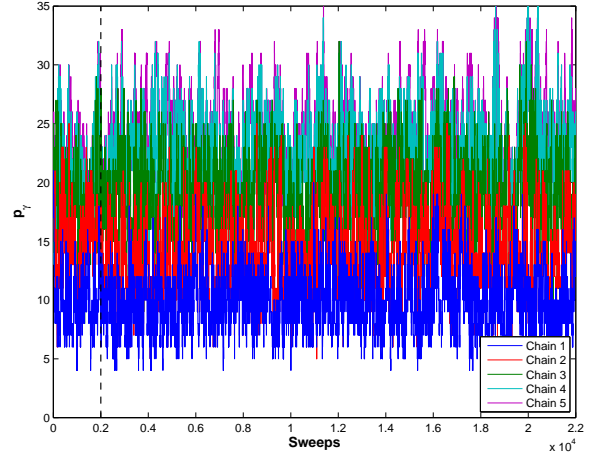}}\\
\subfigure[]{\includegraphics[totalheight=0.25\textheight]{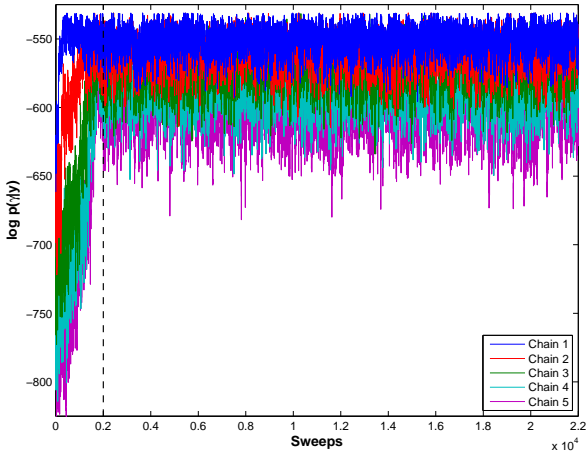}}
\subfigure[]{\includegraphics[totalheight=0.25\textheight]{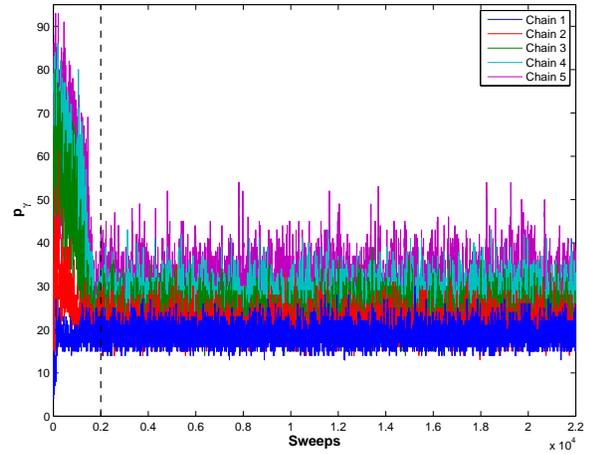}}
\caption{For ESS$i$ with $\tau=1$: (a) trace plot of the log
posterior probability, $\log p\left( \gamma \left\vert y\right.
\right) $, and (b) model size, $p_{\gamma } $, across sweeps for one
replicate of Ex1 with $E\left( p_{\gamma }\right) =20$, top panels
and Ex4, bottom panels. Vertical dashed lines indicate the end of
the burn-in.} \label{Fig_S3}
\end{center}
\end{figure}

In this Section we report some stylised facts about the performance
of the ESS${i}$ with $\tau$ fixed at $1$. Figure \ref{Fig_S3}, top
panels, shows for one of the replicates of Ex1, the overall mixing
properties of ESS$i$. As expected, the chains attached to higher
temperatures shows more variability. Albeit the convergence is
reached in the product space $\prod\nolimits_{l=1}^{L}\left[ p\left(
\gamma _{l}\left\vert y\right. \right) \right] ^{1/t_{l}}$, by
visual inspection each chain \emph{marginally} reaches its
\emph{equilibrium} with respect to the others; moreover, thanks to
the automatic tuning of the temperature placement during the
burn-in, the distributions of their log posterior probabilities
overlap nicely, allowing effective exchange of information between
the chains. Figure \ref{Fig_S3}, bottom panels, shows the trace plot
of the log posterior and the model size for a replicate of Ex4. We
can see that also in the case $p>n$, the chains mix and overlap well
with no gaps between them, the automatic tuning of the temperature
ladder being able to improve drastically the performance of the
algorithm.

This effective exchange of information is demonstrated in Table
\ref{Table_S3} which shows good overall acceptance rates for the
collection of moves that we have implemented. The dimension of the
problem does not seem to affect the acceptance rate of the (delayed
rejection) exchange operator which stays very stable and close to
the target: for instance in Ex4 ($p=300$) and Ex6 ($p=775$) the mean
and standard deviation of the acceptance rate are $0.517$ ($0.105$)
and $0.497$ ($0.072$) while in Ex5 ($p=1,000$) we have $0.505$
($0.013$): the higher variability in Ex4 being related to the model
size $p_{\gamma}$.

With regards to the crossover operators, again we observe stability
across all the examples. Moreover, in contrast to Jasra \emph{et
al.} (2007), when $p>n$, the crossover average acceptance rate
across the five chains is quite stable between $0.147$, Ex4, and
$0.193$, Ex6 (with the lower value in Ex4 here again due to
$p_{\gamma}$): within our limited experiments, we believe that the
good performance of crossover operator is related to the selection
operator and the new block crossover, see Section \ref{EMC sampler}.

Some finer tuning of the temperature ladder could still be performed
as there seems to be an indication that fewer global moves are
accepted with the higher temperature chain, see Table
\ref{Table_S4}, where swapping probabilities for each chain are
indicated. Note that the observed frequency of successful swaps is
not far from the case where adjacent chains are selected to swap at
random with equal probability. Other measures of overlapping between
chains (Liang and Wong, 2000; Iba 2001), based on a suitable index
of variation of $f\left( \gamma \right) =\log p\left( y\left\vert
\gamma \right. \right) +\log p\left( \gamma \right) $ across sweeps,
confirm the good performance of ESS$i$. Again some instability is
present in the high temperature chains, see in Table \ref{Table_S4}
the overlapping index between chains $3,4$ and $4,5$ in Example 3 to
6.


In Ex1, we also investigate the influence of different values of the
prior mean of the model size. We found that the average (standard
deviation in brackets) acceptance rate across replicates for the
delayed rejection exchange operator ranges from $0.493$ ($0.043$) to
0.500 (0.040) for different values of the prior mean on the model
size, while the acceptance rate for the crossover operator ranges
from $0.249$ ($0.021$) to $0.271$ ($0.036$). This strong stability
is not surprising because the automatic tuning modifies the
temperature ladder in order to compensate for $E\left( p_{\gamma
}\right) $. Finally we notice that the acceptance rates for the
local move, when $n>p$, increases with higher values of the prior
mean model size, showing that locally the algorithm moves more
freely with $E\left( p_{\gamma }\right) =20$ than with $E\left(
p_{\gamma }\right) =5$.

\subsection{Performance of ESS$i$ and comparison with SSS} \label{Performance_comparison}

\subsubsection*{Performance of ESS$i$}
We conclude this Section by discussing in details the overall
performance of ESS$i$ with respect to the selection of the true
simulated effects. As a first measure of performance,  we report for
all the simulated examples  the marginal posterior probability of
inclusion as described in G\&McC (1997) and Hans \emph{et al.}
(2007). In the following, for ease of notation, we drop the chain
subscript index and we exclusively refer to the first chain. To be
precise, we evaluate the marginal posterior probability of inclusion
as:
\begin{equation}
p\left( \gamma _{j}=1\left\vert y\right. \right) \simeq
C^{-1}\sum_{t=1,\ldots ,T}1_{\left( \gamma_{j} ^{\left( t\right)
}=1\right) }\left( \gamma \right) p\left( y\left\vert \gamma
^{\left( t\right) }\right. \right) p\left( \gamma ^{\left( t\right)
}\right) \label{L35}
\end{equation}
with $C=\sum_{t=1,\ldots ,T}p\left( y\left\vert \gamma ^{\left(
t\right) }\right. \right) p\left( \gamma ^{\left( t\right) }\right)
$ and $T$ the number of sweeps after the burn-in. The posterior
model size is similarly defined, $p\left( p_{\gamma }\left\vert
y\right. \right) \simeq C^{-1}\sum_{t=1,\ldots ,T}1_{\left(
\left\vert \gamma ^{\left( t\right) }\right\vert =p_{\gamma }\right)
}\left( \gamma \right) p\left( y\left\vert \gamma ^{\left( t\right)
}\right. \right) p\left( \gamma ^{\left( t\right) }\right) $, with
$C$ as before. Besides plotting the marginal posterior inclusion
probability (\ref{L35}) averaged across sweeps and replicates for
our simulated examples, we will also compute the interquartile range
of (\ref{L35}) across replicates as a measure of variability.

In order to thoroughly compare the proposed ESS algorithm to SSS
(Hans \emph{et al.}, 2007), we present  also some other measures of
performance based on $p\left( \gamma \left\vert y\right. \right) $
and $R_{\gamma }^{2}$ : first we rank  $p\left( \gamma \left\vert
y\right. \right) $ in decreasing order and record the indicator
$\gamma $ that corresponds to the maximum and $1,000$ largest
$p\left( \gamma \left\vert y\right. \right) $ (after burn-in). Given
the above set of latent binary vectors, we then compute the
corresponding $R_{\gamma }^{2}$ leading to \lq\lq $R_{\gamma }^{2}$:
$\max p\left( \gamma \left\vert y\right. \right) $\rq\rq\ as well as
the mean $R_{\gamma }^{2}$ over the $1,000$ largest $p\left( \gamma
\left\vert y\right. \right) $, \lq\lq $\overline{R_{\gamma }^{2}}$:
$1,000$ largest $p\left( \gamma \left\vert y\right. \right) $\rq\rq,
both quantities averaged across replicates. Moreover the actual
ability of the algorithm to reach regions of high posterior
probability and persist on them is monitored: given the sequence of
the $1,000$ best $\gamma$s (based on $p\left( \gamma \left\vert
y\right. \right) $), the standard deviation of the corresponding
$R_{\gamma }^{2}$s shows how stable is the searching strategy at
least for the top ranked (not unique) posterior probabilities:
averaging over the replicates, it provides an heuristic measures of
\lq\lq stability\rq\rq\ of the algorithm. Finally we report the
average computational time (in minutes) across replicates of ESS$i$
written in Matlab code and run on a 2MHz CPU with 1.5 Gb RAM desktop
computer and of SSS version 2.0 on the same computer.

\subsubsection*{Comparison with SSS}
\begin{figure}
\subfigure[]{\includegraphics[totalheight=0.25\textheight]{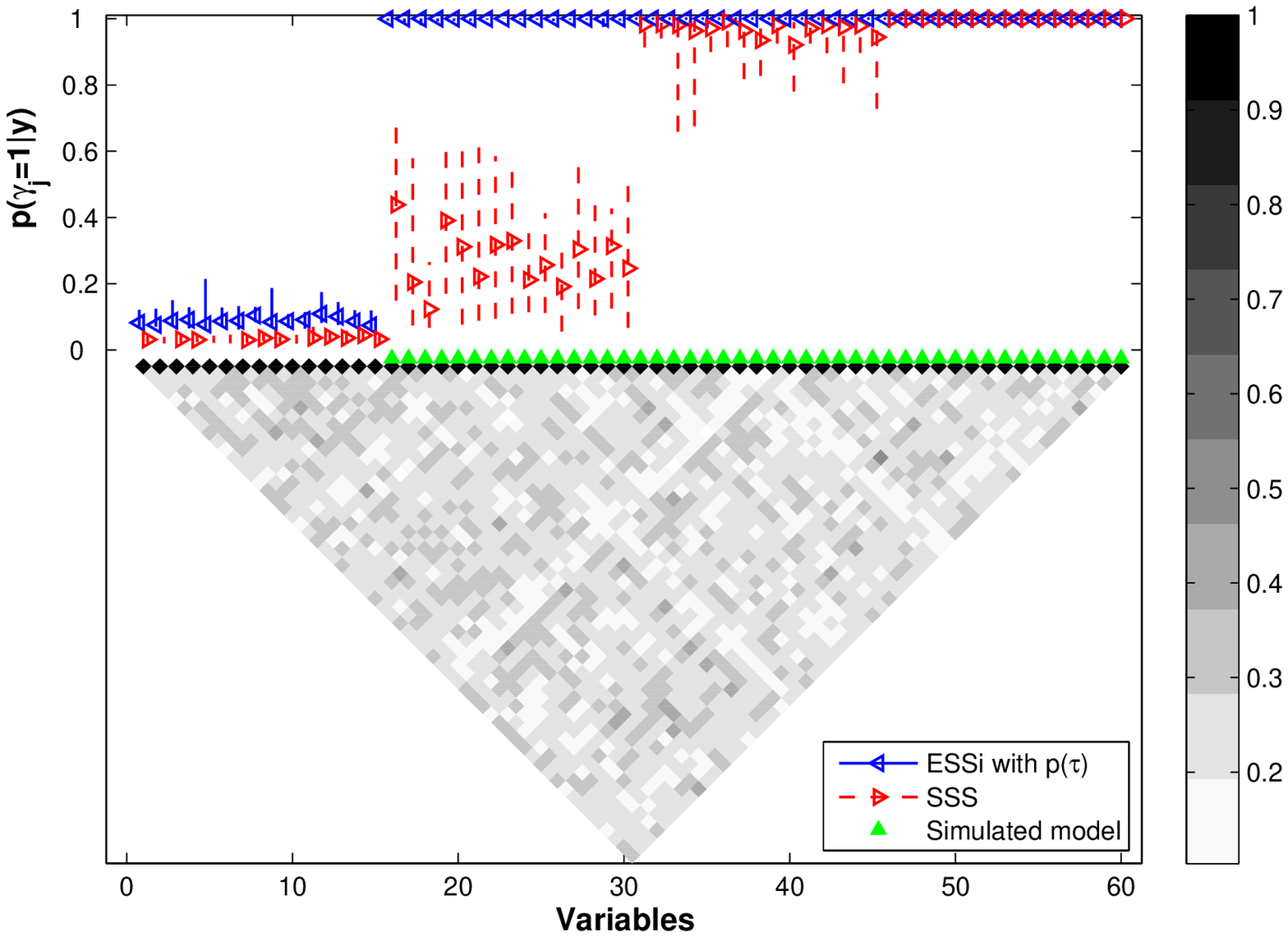}}
\subfigure[]{\includegraphics[totalheight=0.25\textheight]{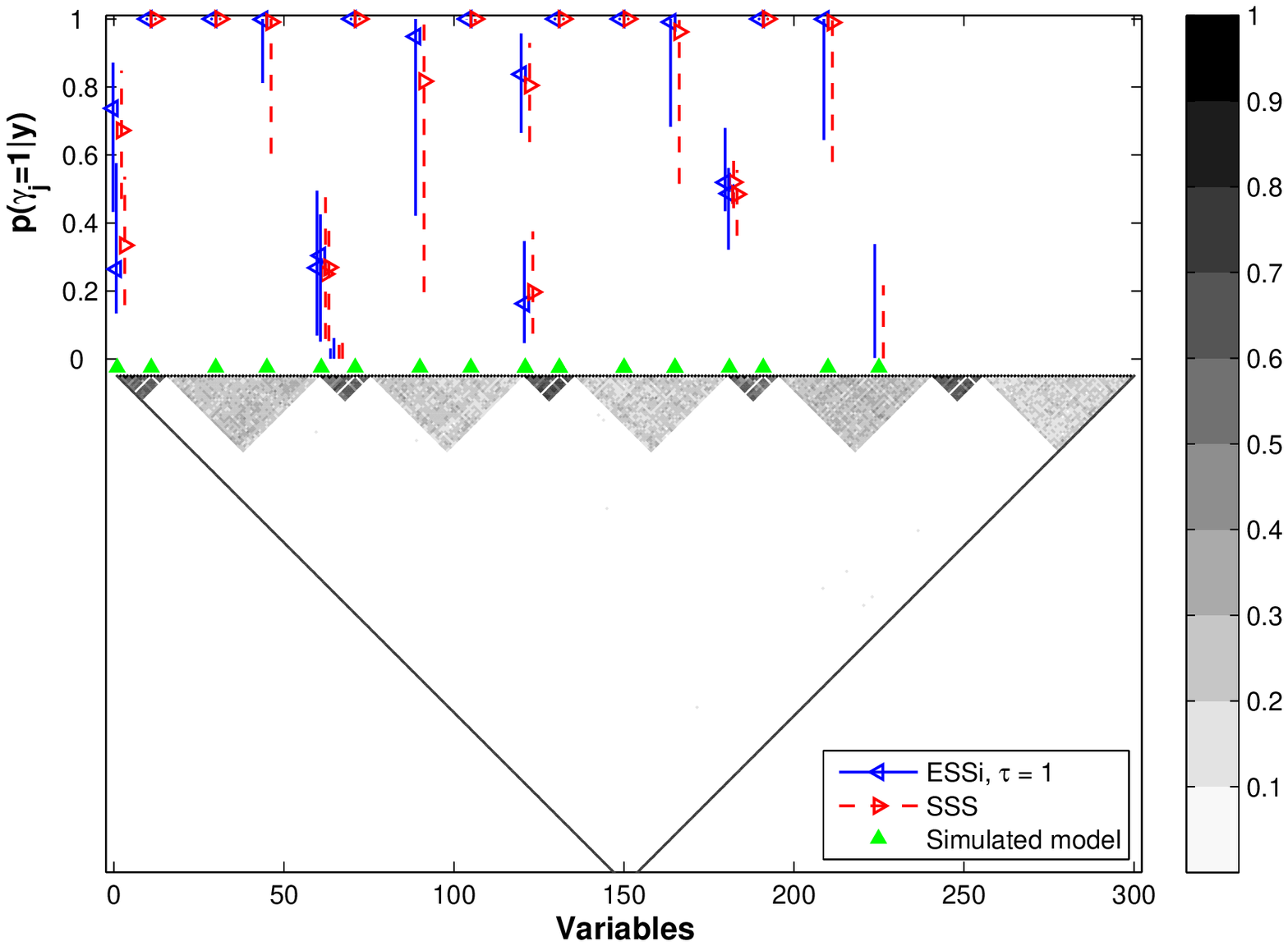}}\\
\subfigure[]{\includegraphics[totalheight=0.25\textheight]{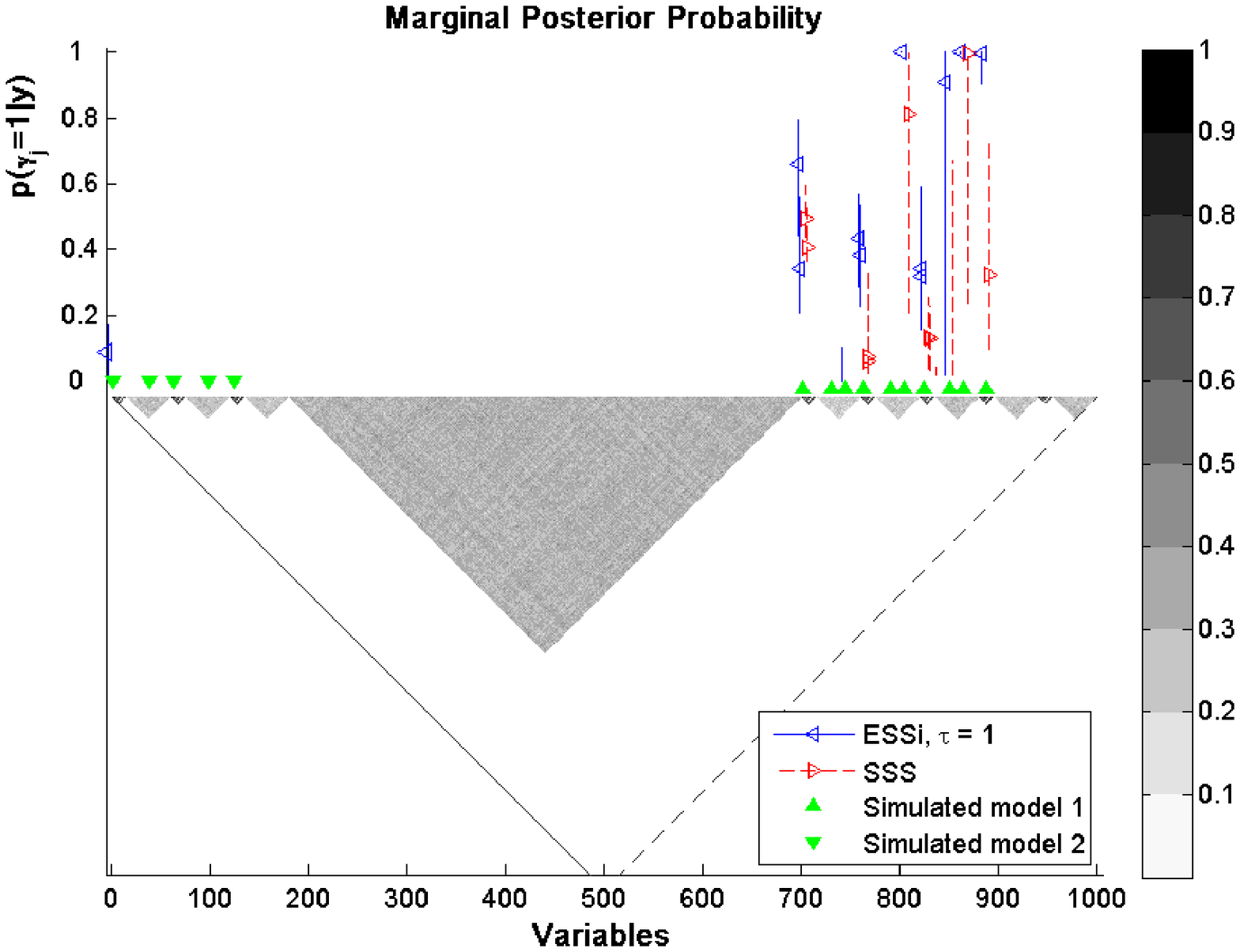}}
\subfigure[]{\includegraphics[totalheight=0.25\textheight]{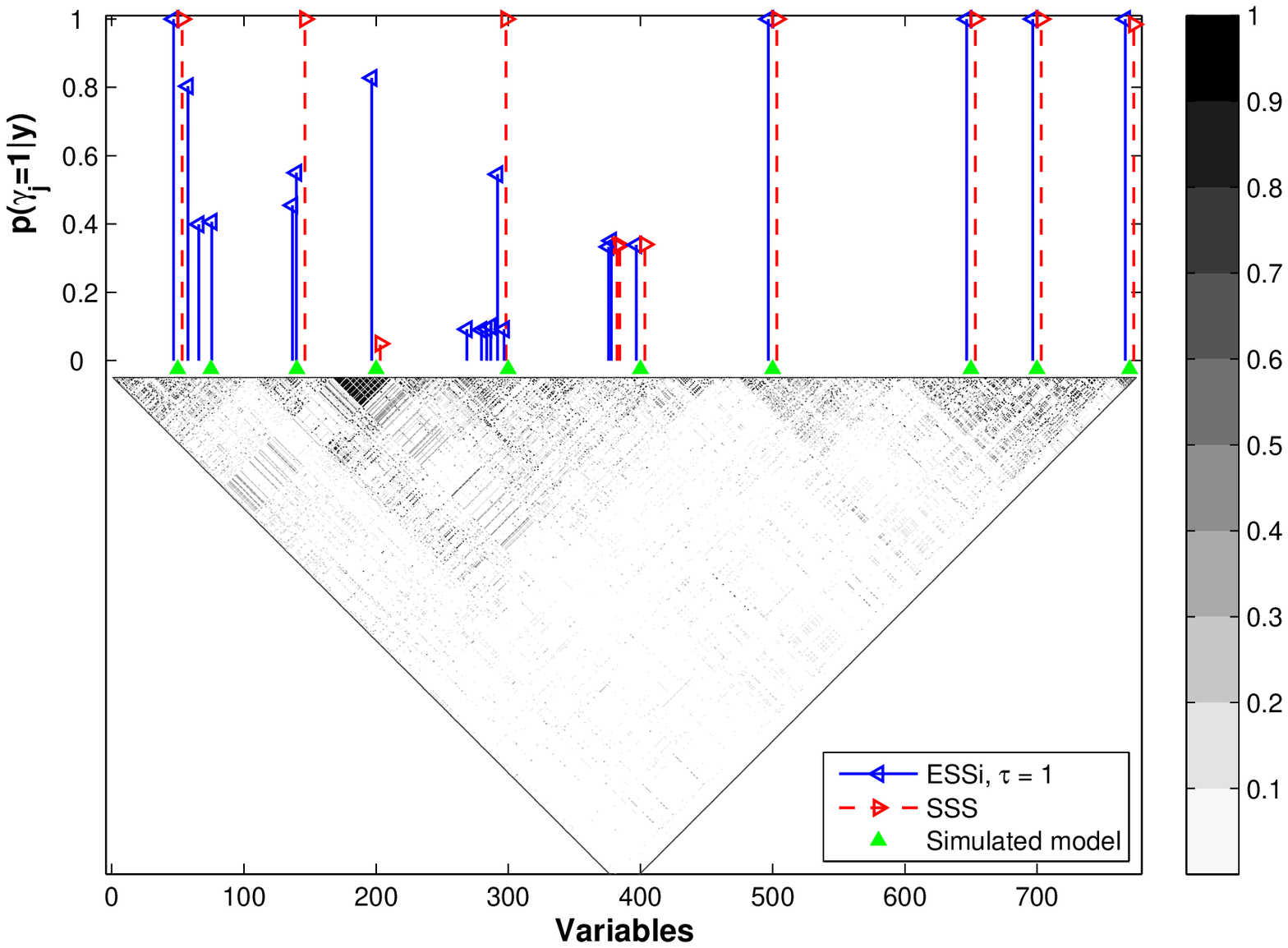}}
\caption{Median and interquartile range of the marginal posterior
probability of inclusion (\ref{L35}) for Ex3, (a), Ex4, (b) and Ex5,
(c), across replicates. Each graph is constructed as follows: bottom
part, pairwise squared correlation $\rho^2 \left( X_{j},X_{j^{\prime
}}\right) $, $j=1,\ldots ,p$, between predictors for one selected
replicate, grey scale indicates different values of squared
correlation; blue left and red right triangles, median of $p\left(
\gamma _{j}=1\left\vert y\right. \right)$ across replicates for
ESS$i$ with $\tau=1$ and SSS respectively; vertical blue solid lines
and vertical red dashed lines, interquartile range of $p\left(
\gamma _{j}=1\left\vert y\right. \right)$ across replicates for
ESS$i$ and SSS respectively; upper and lower green triangles,
simulated models. Selected replicate of Ex6, (d), shows marginal
posterior probability of inclusion (blue left and red right
triangles for ESS$i$ $\tau=1$ and SSS respectively). Marginal
posterior probability of inclusion lower than $0.025$ not shown.}
\label{Fig_S4}
\end{figure}

Figure \ref{Fig_S4} presents the marginal posterior probability of
inclusion for ESS$i$ with $\tau=1$ averaged across replicates and,
as a measure of variability, the interquartile range, blue left
triangles and vertical blue solid line respectively. In general the
covariates with non-zero effects have high marginal posterior
probability of inclusion in all the examples: for example in Ex3,
Figure \ref{Fig_S4} (a), the proposed ESS$i$ algorithm, blue left
triangle, is able to perfectly select the last $45$ covariates,
while the first $15$, which do not contribute to $y$, receive small
marginal posterior probability. It is interesting to note that this
group of covariates, $\left( \beta _{1},\ldots ,\beta _{15}\right)
=\left( 0,\ldots ,0\right) $, although correctly recognised having
no influence on $y$, show some variability across replicates,
vertical blue solid line: however, this is not surprising since
independent priors are less suitable in situations where all the
covariates are mildly-strongly correlated as in this simulated
example. On the other hand the second set of covariates with small
effects, $\left( \beta _{16},\ldots ,\beta _{30}\right) =\left(
1,\ldots ,1\right) $, are univocally detected. The ability of ESS$i$
to select variables with small effects is also evident in Ex6,
Figure \ref{Fig_S4} (d), where the two smallest coefficients,
$\beta_{2}=0.112$ and $\beta_{10}=0.950$ (the second and last
respectively from left to right), receive from high to very high
marginal posterior probability (and similarly for the other
replicates, data not shown). In some cases however, some covariates
attached with small effects are missed (e.g. Ex4, Figure
\ref{Fig_S4} (b), the last simulated effect which is also the
smallest, $\beta_{16}=0.5$, is not detected). In this situation
however the vertical blue solid line indicates that for some
replicates, ESS$i$ is able to assign small values of the marginal
posterior probability giving evidence that ESS$i$ fully explore the
whole space of models.

Superimposed on all pictures of Figure \ref{Table_S4} are the median
and interquartile range across replicates of $p\left( \gamma
_{j}=1\left\vert y\right. \right) $, $j=1,\ldots ,p$, for SSS, red
right triangles and vertical red dashed line respectively. We see
that there is good agreement between the two algorithms in general,
with in addition evidence that ESS$i$ is able to explore more fully
the model space and in particular to find small effects, leading to
a posterior model size that is close to the true one. For instance
in Ex3, Figure \ref{Fig_S4} (a), where the last $30$ covariates
accounts for most of $R_{\gamma }^{2}$, SSS has difficulty to detect
$\left( \beta _{16},\ldots ,\beta _{30}\right) $, while in Ex6, it
misses $\beta_{2}=0.112$, the smallest effect, and surprisingly also
$\beta_{4} = -2.595$ assigning a very small marginal posterior
probability (and in general for the small effects in most
replicates, data not shown). However the most marked difference
between ESS$i$ and SSS is present in Ex5: as for ESS$i$, SSS misses
three effects of \lq\lq model 1\rq\rq\, but in addition $\beta_{4} =
1$, $\beta_{7} = -1$ and $\beta_{8} = 1.5$ receive also very low
marginal posterior probability, red right triangle, with high
variability across replicates, vertical red dashed line. Moreover on
the extreme left, as noted before, ESS$i$ is able to capture the
biggest coefficient of \lq\lq model 2\rq\rq\, while SSS misses
completely all contaminated effects. No noticeable differences
between ESS$i$ and SSS are present in Ex1 and Ex2 for the marginal
posterior probability, while in Ex4, SSS shows more variability in
$p\left( \gamma _{j}=1\left\vert y\right. \right) $ (red dashed
vertical lines compared to blue solid vertical lines) for some
covariates that do receive the highest marginal posterior
probability.

In contrast to the differences in the marginal posterior probability
of inclusion, there is general agreement between the two algorithms
with respect to some measures of goodness of fit and stability, see
Table \ref{Table_S5}. Again, not surprisingly, the main difference
is seen in Ex5 where ESS$i$ with $\tau=1$ reaches a better
$R_{\gamma }^{2}$ both for the maximum and the $1,000$ largest
$p\left( \gamma \left\vert y\right. \right) $. SSS shows more
stability in all examples, but the last: this was somehow expected
since one key features of SSS in its ability to move quickly towards
the right model and to persist on it (Hans \emph{et al.}, 2007), but
a drawback of this is its difficulty to explore far apart models
with competing $R_{\gamma }^{2}$ as in Ex5. Note that ESS$i$ shows a
small improvement of $R_{\gamma }^{2}$ in all the simulated
examples. This is related to the ability of ESS$i$ to pick up some
of the small effects that are missed by SSS, see Figure
\ref{Fig_S4}. Finally ESS$i$ shows a remarkable superiority in terms
of computational time especially when the simulated (and estimated)
$p_{\gamma }$ is large (in other simulated examples, data not shown,
we found this is always true when $p_{\gamma }\gtrsim 10$): the
explanation lies in the number of different models SSS and ESS$i$
evaluate at each sweep. Indeed, SSS evaluates $p+p_{\gamma }\left(
p-p_{\gamma }\right) $, where $p_{\gamma }$ is the size of the
current model, while ESS$i$ theoretically analyses an equally large
number of models, $pL$, but, when $p>n$, the actual number of models
evaluated is drastically reduced thanks to our FSMH sampler. In only
one case SSS beats ESS$i$ in term of computational time (Ex5), but
in this instance SSS clearly underestimates the simulated model and
hence performs less evaluations than would be necessary to explore
faithfully the model space. In conclusion, we see that the rich
porfolio of moves and the use of parallel chains makes ESS robust
for tackling complex covariate space as well as competitive against
a state of the art search algorithm.

\end{appendix}


\renewcommand{\baselinestretch}{1.5}


\section*{References}

\hangindent=1.5em \hangafter=1 \noindent Abramowitz, M. and Stegun,
I. (1970). \emph{Handbook of Mathematical Functions}. New York:
Dover Publications, Inc.

\hangindent=1em \hangafter=1 \noindent Altshuler, D., Brooks, L.D.,
Chakravarti, A., Collins, F.S., Daly, M.D. and Donnelly, P. (2005).
A haplotype map of the human genome. \emph{Nature}, \textbf{437},
1299-1320.

\hangindent=1em \hangafter=1 \noindent Bae, N. and Mallick, B.K.
(2004). Gene selection using a two-level hierarchical Bayesian
model. \emph{Bioinformatics}, \textbf{20}, 3423-3430.




\hangindent=1em \hangafter=1 \noindent Brown, P.J., Vannucci, M. and
Fearn, T. (1998). Multivariate Bayesian variable selection and
prediction. \emph{J. R. Statist. Soc. B}, \textbf{60}, 627-641.


\hangindent=1em \hangafter=1 \noindent Calvo, F. (2005) All-exchange
parallel tempering. \emph{J. Chem. Phys.}, \textbf{123}, 1-7.


\hangindent=1em \hangafter=1 \noindent Chipman, H. (1996). Bayesian
variable selection with related predictors. \emph{Canad. J.
Statist.}, \textbf{24}, 17-36.

\hangindent=1em \hangafter=1 \noindent Chipman, H., George, E.I. and
McCulloch, R.E. (2001). The practical implementation of Bayesian
model selection (with discussion). In \emph{Model Selection} (P.
Lahiri, ed), 66-134. IMS: Beachwood, OH.

\hangindent=1em \hangafter=1 \noindent Clyde, M. and George, E. I.
(2004). Model uncertainty. \emph{Statist. Sci.}, \textbf{19}, 81-94.


\hangindent=1em \hangafter=1 \noindent Cui, W. and George, E.I.
(2008). Empirical Bayes vs fully Bayes variable selection. \emph{J.
Stat. Plan. Inf.}, \textbf{138}, 888-900.

\hangindent=1em \hangafter=1 \noindent Dellaportas, P., Forster, J.
and Ntzoufras, I. (2002). On Bayesian model and variable selection
using MCMC. \emph{Statist. Comp.}, \textbf{12}, 27-36.



\hangindent=1em \hangafter=1 \noindent Fern\'{a}ndez, C., Ley, E.
and Steel, M.F.J. (2001). Benchmark priors for Bayesian model
averaging. \emph{J. Econometrics}, \textbf{75}, 317-343.


\hangindent=1em \hangafter=1 \noindent George, E.I. and McCulloch,
R.E. (1993). Variable selection via Gibbs sampling. \emph{J. Am.
Statist. Assoc.}, \textbf{88}, 881-889.

\hangindent=1em \hangafter=1 \noindent George, E.I. and McCulloch,
R.E. (1997). Approaches for Bayesian variable selection. \emph{Stat.
Sinica}, \textbf{7}, 339-373.

\hangindent=1em \hangafter=1 \noindent Geweke, J. (1996). Variable
selection and model comparison in regression. In \emph{Bayesian
Statistics 5, Proc. 5th Int. Meeting} (J.M. Bernardo, J.O. Berger,
A.P. Dawid and A.F.M. Smith, eds), 609-20. Claredon Press: Oxford,
UK.

\hangindent=1em \hangafter=1 \noindent Goswami, G. and Liu, J.S.
(2007). On learning strategies for evolutionary Monte Carlo.
\emph{Statist. Comp.}, \textbf{17}, 23-38.

\hangindent=1em \hangafter=1 \noindent Gramacy, R.B, J. Samworth,
R.J. and King, R. (2007). Importance Tempering. Tech. rep. Available
at: \texttt{http://arxiv.org/abs/0707.4242}

\hangindent=1em \hangafter=1 \noindent Green, P. and Mira, A.
(2001). Delayed rejection in reversible jump Metropolis-Hastings.
\emph{Biometrika}, \textbf{88}, 1035-1053.

\hangindent=1em \hangafter=1 \noindent Iba, Y. (2001). Extended
Ensemble Monte Carlo. \emph{Int. J. Mod. Phys., C}, \textbf{12},
623-656.

\hangindent=1em \hangafter=1 \noindent Hans, C., Dobra, A. and West,
M. (2007). Shotgun Stochastic Search for \lq\lq large $p$\rq\rq\
regression. \emph{J. Am. Statist. Assoc.}, \textbf{102}, 507-517.

\hangindent=1em \hangafter=1 \noindent Hubner, N. \emph{et al.}
(2005). Integrated transcriptional profiling and linkage analysis
for identification of genes underlying disease. \emph{Nat. Genet.},
\textbf{37}, 243-253.


\hangindent=1em \hangafter=1 \noindent Kohn, R., Smith, M. and Chan,
D. (2001). Nonparametric regression using linear combinations of
basis functions. \emph{Statist. Comp.}, \textbf{11}, 313-322.

\hangindent=1em \hangafter=1 \noindent Jasra, A., Stephens, D.A. and
Holmes, C. (2007). Population-based reversible jump Markov chain
Monte Carlo. \emph{Biometrika}, \textbf{94}, 787-807.

\hangindent=1em \hangafter=1 \noindent Liang, F., Paulo, R., Molina,
G., Clyde, M.A. and Berger, J.O. (2008). Mixtures of $g$-priors for
Bayesian variable selection. \emph{J. Am. Statist. Assoc.},
\textbf{481}, 410-423.

\hangindent=1em \hangafter=1 \noindent Liang, F. and Wong, W.H.
(2000). Evolutionary Monte Carlo: application to $C_{p}$ model
sampling and change point problem. \emph{Stat. Sinica}, \textbf{10},
317-342.


\hangindent=1em \hangafter=1 \noindent Liu, J.S. (2001). \emph{
Monte Carlo strategies in scientific computations}. Springer: New
York.

\hangindent=1em \hangafter=1 \noindent Madigan, D. and York, J.
(1995). Bayesian graphical models for discrete data. \emph{Int.
Statist. Rev.}, \textbf{63}, 215-232.


\hangindent=1em \hangafter=1 \noindent Natarajan, R. and McCulloch.
(1998). Gibbs sampling with diffuse proper priors: a valid approach
to data-driven inference?, \emph{J. Comp. Graph. Statist.},
\textbf{7}, 267-277.

\hangindent=1em \hangafter=1 \noindent Nott, D.J. and Green, P.J.
(2004). Bayesian variable selection and the Swedsen-Wang algorithm.
\emph{J. Comp. Graph. Statist.}, \textbf{13}, 141-157.

\hangindent=1em \hangafter=1 \noindent Roberts, G.O. and Rosenthal,
J.S. (2008). Example of adaptive MCMC. Tech. rep.  Available at:
\texttt{http://www.probability.ca/jeff/research.html}



\hangindent=1em \hangafter=1 \noindent Tierney, L. and Kadane, J.B.
(1986). Accurate approximations for posterior moments and marginal
densities. \emph{J. Am. Statist. Assoc.}, \textbf{81}, 82-86.

\hangindent=1em \hangafter=1 \noindent Wilson, M.A., Iversen, E.S.,
Clyde, M.A., Schmidler, S.C. and Shildkraut, J.M. (2009). Bayesian
model search and multilevel inference for SNP association studies.
Tech. rep. Available at: \\ \texttt{http://arxiv.org/abs/0908.1144}

\hangindent=1em \hangafter=1 \noindent Zellner, A. (1986). On
assessing prior distributions and Bayesian regression analysis with
$g$-prior distributions. In \emph{Bayesian Inference and Decision
Techniques-Essays in Honour of Bruno de Finetti} (P.K. Goel and A.
Zellner, eds), 233-243. Amsterdam: North-Holland.

\hangindent=1em \hangafter=1 \noindent Zellner, A. and Siow, A.
(1980). Posterior odds ratios for selected regression hypotheses. In
\emph{Bayesian Statistics, Proc. 1st Int. Meeting} (J.M. Bernardo,
M.H. De Groot, D.V. Lindley and A.F.M. Smith, eds), 585-603.
Valencia: University Press.


\renewcommand{\baselinestretch}{1.5}



\clearpage

\clearpage

\clearpage

\clearpage

\clearpage

\clearpage

%



\clearpage

\begin{table}
\begin{center} \fbox{
\begin{tabular}{ll|ccccc}
&  & Mode$\left( p_{\gamma }\left\vert y\right. \right) $ & $E\left(
\tau \left\vert y\right. \right) $ & $ \left. R_{\gamma }^{2}\right.
^{\ast } $ & $ \left. \overline{R_{\gamma }^{2}}\right. ^{\ast \ast
} $ & Stability \\ \hline \multirow{2}{*}{eQTL} & ESS$g$, $\tau =29$
& $2$ & $-$ & $0.716$ & $0.704$ & $0.257$ \\ & ESS$g$ with $p\left(
\tau \right) $ & $2$ & $20.576$ & $0.716$
& $0.689$ & $ 0.099$ \\
\cline{1-2} \multirow{2}{*}{mQTL} & ESS$g$, $\tau =50$ & $2$ & $-$ &
$0.843$ & $0.843$ & $\approx 0$ \\ & ESS$g$ with $p\left( \tau
\right) $ & $2$ & $63.577$ & $0.843$ &
$0.843$ & $\approx 0$ \\
\hline &  & Crossover & $ \text{DR Exchange} $ & $ \text{ALL
Exchange} $ & $ \text{Acc. rate }\tau$ & Time (min.) \\ \hline
\multirow{2}{*}{eQTL} & ESS$g$, $\tau =29$ & $0.214$ & $0.534$ &
$0.671$ & $-$ & $28$ \\ & ESS$g$ with $p\left( \tau \right) $ &
$0.243$ & $0.585$ &
$0.711$ & $0.438$ & $30$ \\
\cline{1-2} \multirow{2}{*}{mQTL} & ESS$g$, $\tau =50$ & $0.214$ &
$0.514$ & $0.669$ & $-$ & $302$ \\ & ESS$g$ with $p\left( \tau
\right) $ & $0.226$ & $0.571$ & $0.717$
& $0.434$ & $309$ \\
\end{tabular}}
\caption{Performance of ESS$g$ with and without the hyperprior on
$\tau$ for the first real data example, eQTL analysis, and second
example, mQTL analysis. $\left. R_{\gamma }^{2}\right. ^{\ast }$ and
$\left. \overline{R_{\gamma }^{2}}\right. ^{\ast \ast }$ correspond
to \lq\lq $R_{\gamma }^{2}$: $\max p\left( \gamma \left\vert
y\right. \right) $\rq\rq\ and \lq\lq $\overline{R_{\gamma }^{2}}$:
$1,000$ largest $p\left( \gamma \left\vert y\right. \right) $\rq\rq\
respectively. The former indicates the coefficient of determination
for the (first chain) best model visited according to the posterior
probability $p\left( \gamma \left\vert y\right. \right) $, while the
latter shows the average $R_{\gamma }^{2}$ for the (first chain) top
$1,000$ (not unique) visited models ranked by the posterior
probability. \lq\lq Stability\rq\rq\ is defined as the standard
deviation of $R_{\gamma }^{2}$ for the (first chain) top $1,000$
(not unique) visited models (smaller values indicate better
performance of the algorithm). In the bottom part of the Table,
acceptance rate for specific moves are given. \lq\lq DR
Exchange\rq\rq\ and \lq\lq ALL Exchange\rq\rq\ stands for \lq\lq
delayed rejection exchange\rq\rq\ and \lq\lq all-exchange\rq\rq\
move respectively. } \label{Table_T1}
\end{center}
\end{table}

\clearpage

\begin{table}
\begin{center} \fbox{
\begin{tabular}{c|c|cc}
& Version of ESS$g$ & $\tau $ & $p\left( \tau \right) $ \\ \hline
\multicolumn{1}{l|}{\multirow{2}{*}{Experiment (i)}} &
\multicolumn{1}{|l|}{ESS$g$ with only
FSMH} & \multicolumn{1}{|l}{$68\%$} & \multicolumn{1}{l}{$88\%$} \\
\multicolumn{1}{l|}{} & \multicolumn{1}{|l|}{ESS$g$ with only
MC$^{3}$} & \multicolumn{1}{|l}{$28\%$} & \multicolumn{1}{l}{$40\%$}
\\ \hline \multicolumn{1}{l|}{\multirow{4}{*}{Experiment (ii)}} & \multicolumn{1}{|l|}{ESS$g$ with
only 1-point
crossover} & \multicolumn{1}{|l}{$64\%$} & \multicolumn{1}{l}{$80\%$} \\
\multicolumn{1}{l|}{} & \multicolumn{1}{|l|}{ESS$g$ with only block
crossover} & \multicolumn{1}{|l}{$80\%$} &
\multicolumn{1}{l}{$84\%$}
\\
\multicolumn{1}{l|}{} & \multicolumn{1}{|l|}{ESS$g$ with only
uniform
crossover} & \multicolumn{1}{|l}{$60\%$} & \multicolumn{1}{l}{$84\%$} \\
\multicolumn{1}{l|}{} & \multicolumn{1}{|l|}{ESS$g$ with only
adaptive crossover} & \multicolumn{1}{|l}{$60\%$} &
\multicolumn{1}{l}{$76\%$}
\end{tabular}}
\caption{Proportion of times different versions of ESS$g$ reach the
same top visited model in the eQTL real data set with or without an
hyperprior on $\tau$ in $25$ replicates of the analysis.}
\label{Table_S1}
\end{center}
\end{table}

\begin{table}
\begin{center} \fbox{
\begin{tabular}{c|c|cc}
& Version of ESS$g$ & $\tau $ & $p\left( \tau \right) $ \\ \hline
\multicolumn{1}{l|}{\multirow{4}{*}{Experiment (ii)}} &
\multicolumn{1}{|l|}{ESS$g$ with only 1-point crossover} &
\multicolumn{1}{|l}{$0.303$} & \multicolumn{1}{l}{$0.335
$} \\
\multicolumn{1}{l|}{} & \multicolumn{1}{|l|}{ESS$g$ with only block
crossover
} & \multicolumn{1}{|l}{$0.482$} & \multicolumn{1}{l}{$0.501$} \\
\multicolumn{1}{l|}{} & \multicolumn{1}{|l|}{ESS$g$ with only
uniform
crossover} & \multicolumn{1}{|l}{$0.026$} & \multicolumn{1}{l}{$0.042$} \\
\multicolumn{1}{l|}{} & \multicolumn{1}{|l|}{ESS$g$ with only
adaptive crossover} & \multicolumn{1}{|l}{$0$} &
\multicolumn{1}{l}{$0.013$}
\end{tabular}}
\caption{Average acceptance rate of the crossover operator for
different versions of ESS$g$ in $25$ replicates of the analysis of
the first real data example with or without an hyperprior on
$\tau$.} \label{Table_S2}
\end{center}
\end{table}

\clearpage

\begin{table}
\begin{center} \fbox{
\begin{tabular}{l|cccccccc}{\rule[0mm]{0mm}{5mm}} &  & Ex1 &  & Ex2 & Ex3 &
Ex4 & Ex5 & Ex6 \\ \cline{2-4} {\rule[0mm]{0mm}{5mm}}
\multirow{2}{*}{$
\begin{array}{c}
{\small n} \\
{\small p}
\end{array}
$} & & $120$ &
& $300$ & $120$ & $120$ & $200$ & $120$ \\
{\rule[0mm]{0mm}{5mm}} & & $60$ &  & $30$ &
$60$ & $300$ & $1,000$ & $775$ \\
{\rule[0mm]{0mm}{5mm}} $E\left( p_{\gamma }\right) $ & $5$ &
$10$ & $20$ & $5$ & $5$ & $5$ & $5$ & $ 5$ \\
\cline{1-9}{\rule[0mm]{0mm}{5mm}} \multirow{2}{*}{Add/delete} &
$0.036$ & $0.054$ & $0.098$ & $0.066$ & $0.086$ & - &
- & - \\
& $\left( 0.016\right) $ & $\left( 0.017\right) $ & $\left(
0.023\right) $ &
$\left( 0.020\right) $ & $\left( 0.031\right) $ & - & - & - \\
{\rule[0mm]{0mm}{5mm}} \multirow{2}{*}{Swap} & $0.063$ & $0.100$ &
$0.165$ & $0.070$ & $0.106$ & - & - & -
\\
& $\left( 0.015\right) $ & $\left( 0.019\right) $ & $\left(
0.022\right) $ &
$\left( 0.015\right) $ & $\left( 0.053\right) $ & - & - & - \\
{\rule[0mm]{0mm}{5mm}} \multirow{2}{*}{Crossover} & $0.249$ &
$0.270$ & $0.271$ & $0.157$ & $0.215$ & $0.147$ & $
0.170$ & $0.193$ \\
& $\left( 0.021\right) $ & $\left( 0.029\right) $ & $\left(
0.036\right) $ & $\left( 0.018\right) $ & $\left( 0.022\right) $ &
$\left( 0.028\right) $ & $
\left( 0.023\right) $ & $\left( 0.028\right) $ \\
{\rule[0mm]{0mm}{5mm}} \multirow{2}{*}{DR Exchange} & $0.500$ &
$0.493$ & $0.500$ & $0.582$ & $0.492$ & $0.517$ & $0.505
$ & $0.497$ \\
& $\left( 0.040\right) $ & $\left( 0.043\right) $ & $\left(
0.040\right) $ & $\left( 0.020\right) $ & $\left( 0.071\right) $ &
$\left( 0.105\right) $ & $ \left( 0.013\right) $ & $\left(
0.072\right) $
\end{tabular}}
\caption{Mean and standard deviation in brackets of EMC acceptance
rates across replicates for ESS$i$ with $\tau=1$. \lq\lq DR
Exchange\rq\rq\ stands for \lq\lq delayed rejection exchange\rq\rq.}
\label{Table_S3}
\end{center}
\end{table}

\begin{table}
\begin{center} \fbox{
\begin{tabular}{l|ccccccccc}{\rule[0mm]{0mm}{5mm}}
&  &  & Ex1 &  & Ex2 & Ex3 & Ex4 & Ex5 & Ex6 \\ \cline{3-5}
{\rule[0mm]{0mm}{5mm}}\multirow{2}{*}{$
\begin{array}{c}
{\small n} \\
{\small p}
\end{array}
$} &  &  & $120$ &  & $300$ &
$120$ & $120$ & $200$ & $120$ \\
{\rule[0mm]{0mm}{5mm}} &  &  & $60$ &  & $30$ &
$60$ & $300$ & $1,000$ & $775$ \\
{\rule[0mm]{0mm}{5mm}}$E\left( p_{\gamma }\right) $ &  & $5$ & $10$
& $20$ & $5$ & $5$ & $5$ & $5$ & $5$ \\\hline
Swapping{\rule[6.5mm]{0mm}{5mm}}& \multicolumn{1}{|c}{$
\begin{array}{c}
l=1 \\
l=2 \\
l=3 \\
l=4 \\
l=5
\end{array}
$} & $
\begin{array}{c}
0.157 \\
0.250 \\
0.220 \\
0.240 \\
0.142
\end{array}
$ & $
\begin{array}{c}
0.137 \\
0.232 \\
0.220 \\
0.252 \\
0.160
\end{array}
$ & $
\begin{array}{c}
0.110 \\
0.204 \\
0.223 \\
0.280 \\
0.182
\end{array}
$ & $
\begin{array}{c}
0.065 \\
0.185 \\
0.255 \\
0.293 \\
0.201
\end{array}
$ & $
\begin{array}{c}
0.160 \\
0.271 \\
0.245 \\
0.215 \\
0.110
\end{array}
$ & $
\begin{array}{c}
0.180 \\
0.276 \\
0.223 \\
0.206 \\
0.112
\end{array}
$ & $
\begin{array}{c}
0.201 \\
0.300 \\
0.231 \\
0.182 \\
0.083
\end{array}
$ & $
\begin{array}{c}
0.214 \\
0.316 \\
0.231 \\
0.167 \\
0.070
\end{array}
$ \\ Overlapping &{\rule[5mm]{0mm}{5mm}} $
\begin{array}{c}
l=1,2 \\
l=2,3 \\
l=3,4 \\
l=4,5
\end{array}
$ & $
\begin{array}{c}
1.360 \\
1.570 \\
1.400 \\
1.100
\end{array}
$ & $
\begin{array}{c}
1.600 \\
1.570 \\
1.290 \\
0.992
\end{array}
$ & $
\begin{array}{c}
2.101 \\
1.600 \\
1.050 \\
0.690
\end{array}
$ & $
\begin{array}{c}
2.680 \\
0.870 \\
0.600 \\
1.251
\end{array}
$ & $
\begin{array}{c}
1.350 \\
1.430 \\
2.111 \\
4.131
\end{array}
$ & $
\begin{array}{c}
0.733 \\
1.021 \\
1.329 \\
1.503
\end{array}
$ & $
\begin{array}{c}
0.569 \\
0.913 \\
1.491 \\
2.304
\end{array}
$ & $
\begin{array}{c}
0.526 \\
0.893 \\
1.696 \\
2.499
\end{array}
$
\end{tabular}}
\caption{Swapping probability for ESS$i$ with $\tau=1$ defined as
the observed frequency of successful swaps for each chain (including
delayed rejection exchange and all-exchange operators) averaged
across replicates. Overlapping measure defined as $V\left( f\left(
\gamma _{l} \right) \right) \left( 1/t_{l+1}-1/t_{l}\right) ^{2} $,
Liang and Wong (2000) with $f\left( \gamma_{l} \right) =\log p\left(
y\left\vert \gamma_{l} \right. \right) +\log p\left( \gamma_{l}
\right) $. Target value for consecutive chains is $O\left( 1\right)
$.} \label{Table_S4}
\end{center}
\end{table}

\clearpage

\begin{table}
\begin{center} \fbox{
\begin{tabular}{ll|cccccccc}{\rule[0mm]{0mm}{5mm}}
&  &  & Ex1 &  & Ex2 & Ex3 & Ex4 & Ex5 & Ex6 \\ \cline{3-5} &
{\rule[0mm]{0mm}{5mm}}\multirow{2}{*}{$
\begin{array}{c}
{\small n} \\
{\small p}
\end{array}
$} &  & $120$ &  & $300$ &
$120$ & $120$ & $200$ & $120$ \\
{\rule[0mm]{0mm}{5mm}} &  &  & $60$ &  & $30$ &
$60$ & $300$ & $1,000$ & $775$ \\
& {\rule[0mm]{0mm}{5mm}} $E\left( p_{\gamma }\right) $ & $5$ & $10$
&
$20$ & $5$ & $5$ & $5$ & $5$ & $5$ \\
\hline \multirow{8}{*}{$\begin{array}{c}
\text{ESS}i, \\
\tau \text{=1}
\end{array}$} & \multicolumn{1}{|l|}{\multirow{2}{*}{{$\left. R_{\gamma }^{2}\right.
^{\ast }$}}}{\rule[0mm]{0mm}{5mm}}& $0.864$ & $0.867$ & $0.871$
& $0.975$ & $\approx 1$ & $0.962$ & $0.703$ & $0.997$ \\
& \multicolumn{1}{|l|}{} & $\left( 0.029\right) $ & $\left(
0.027\right) $ & $\left( 0.023\right) $ & $\left( 0.003\right) $ &
$\left( \approx 0\right) $ & $\left( 0.011\right) $ & $\left(
0.043\right) $ & $\left( 0.005\right) $
\\
& \multicolumn{1}{|l|}{\multirow{2}{*}{$\left. \overline{R_{\gamma
}^{2}}\right. ^{\ast \ast }$}}{\rule[0mm]{0mm}{5mm}}& $0.863$ &
$0.866$ & $
0.874$ & $0.975$ & $\approx 1$ & $0.957$ & $0.689$ & $0.997$ \\
& \multicolumn{1}{|l|}{} & $\left( 0.027\right) $ & $\left(
0.026\right) $ & $\left( 0.023\right) $ & $\left( 0.003\right) $ &
$\left( \approx 0\right) $ & $\left( 0.014\right) $ & $\left(
0.048\right) $ & $\left( 0.003\right) $
\\
&
\multicolumn{1}{|l|}{\multirow{2}{*}{Stability}}{\rule[0mm]{0mm}{5mm}}&
$0.003$ & $0.003$ & $0.005$ & $
\approx 0$ & $\left( \approx 0\right) $ & $0.005$ & $0.015$ & $0.002$ \\
& \multicolumn{1}{|l|}{} & $\left( 0.001\right) $ & $\left(
0.002\right) $ & $\left( 0.002\right) $ & $\left( \approx 0\right) $
& $\left( \approx 0\right) $ & $\left( 0.004\right) $ & $\left(
0.007\right) $ & $\left(
0.002\right) $ \\
& \multicolumn{1}{|l|}{\multirow{2}{*}{Time
(min.)}}{\rule[0mm]{0mm}{5mm}}& $6$ & $6$ & $7$ & $16$ & $18$ &
$166$ & $338$
& $202$ \\
& \multicolumn{1}{|l|}{} & $\left( <1\right) $ & $\left( <1\right) $
& $ \left( <1\right) $ & $\left( <1\right) $ & $ \left( 1\right) $ &
$\left( 32\right) $ & $ \left( 43\right) $ & $\left( 40\right) $ \\
\hline \multirow{8}{*}{SSS} &
\multicolumn{1}{|l|}{\multirow{2}{*}{{$\left. R_{\gamma }^{2}\right.
^{\ast }$}}}{\rule[0mm]{0mm}{5mm}}& $0.863$ & $0.867$ & $0.870$
& $0.975$ & $\approx 1$ & $0.956$ & $0.577$ & $0.997$ \\
& \multicolumn{1}{|l|}{} & $\left( 0.027\right) $ & $\left(
0.025\right) $ & $\left( 0.024\right) $ & $\left( 0.003\right) $ &
$\left( \approx 0\right) $ & $\left( 0.016\right) $ & $\left(
0.074\right) $ & $\left( 0.004\right) $
\\
& \multicolumn{1}{|l|}{\multirow{2}{*}{$\left. \overline{R_{\gamma
}^{2}}\right. ^{\ast \ast }$}}{\rule[0mm]{0mm}{5mm}}& $0.863$ &
$0.867$ & $
0.870$ & $0.975$ & $0.999$ & $0.955$ & $0.565$ & $0.996$ \\
& \multicolumn{1}{|l|}{} & $\left( 0.027\right) $ & $\left(
0.025\right) $ & $\left( 0.024\right) $ & $\left( 0.003\right) $ &
$\left( \approx 0\right) $ & $\left( 0.016\right) $ & $\left(
0.078\right) $ & $\left( 0.004\right) $
\\
&
\multicolumn{1}{|l|}{\multirow{2}{*}{Stability}}{\rule[0mm]{0mm}{5mm}}&
$0$ & $0$ & $\approx 0$ & $\approx 0$
& $\approx 0$ & $0.001$ & $0.009$ & $0.004$ \\
& \multicolumn{1}{|l|}{} & $\left( 0\right) $ & $\left( 0\right) $ &
$\left( \approx 0\right) $ & $\left( \approx 0\right) $ & $\left(
\approx 0\right) $ & $\left( 0.002\right) $ & $\left( 0.015\right) $
& $\left( 0.006\right) $
\\
& \multicolumn{1}{|l|}{\multirow{2}{*}{Time
(min.)}}{\rule[0mm]{0mm}{5mm}}& $12$ & $12$ & $13$ & $118$ & $497$ &
$502$ & $
169$ & $549$ \\
& \multicolumn{1}{|l|}{} & $\left( 1\right) $ & $\left( 2\right) $ &
$\left( 2\right) $ & $\left( 26\right) $ & $\left( 75\right) $ &
$\left( 241\right) $ & $\left( 81\right) $ & $\left( 159\right) $
\end{tabular}}
\caption{Comparison between ESS$i$ with $\tau=1$ and SSS for the six
simulated examples. Standard deviation in brackets. $\left.
R_{\gamma }^{2}\right. ^{\ast }$ and $\left. \overline{R_{\gamma
}^{2}}\right. ^{\ast \ast }$ correspond to \lq\lq $R_{\gamma }^{2}$:
$\max p\left( \gamma \left\vert y\right. \right) $\rq\rq\ and \lq\lq
$\overline{R_{\gamma }^{2}}$: $1,000$ largest $p\left( \gamma
\left\vert y\right. \right) $\rq\rq\ respectively.} \label{Table_S5}
\end{center}
\end{table}


\end{document}